\documentclass[a4paper,superscriptaddress,twocolumn,noshowkeys]{revtex4}

\usepackage{graphicx}
\usepackage{amsmath}
\usepackage[hypertexnames=false]{hyperref}
\usepackage{siunitx}  
\usepackage{amsfonts}
\usepackage{hhline}
\usepackage{MnSymbol}
\usepackage{float}

\newcommand{\ud}[1]{{#1^{\dagger}}}
\newcommand{\bra}[1]{\left\langle #1\right|}
\newcommand{\ket}[1]{\left| #1\right\rangle}
\newcommand{\kket}[1]{\|#1\rrangle}

\begin{document}

\title{Ultrafast control of Rabi oscillations in a polariton condensate.}

\author{L. Dominici,$^{1,2\ast}$ D. Colas,$^{3}$ S. Donati,$^{1,2}$ J.~P. Restrepo Cuartas,$^{3}$ M. De Giorgi,$^{1,2}$  D. Ballarini,$^{1,2}$ G. Guirales,$^{4}$\\J. C. L\'opez Carre\~no,$^{3}$ A. Bramati,$^{5}$ G. Gigli,$^{1,2,6}$ E. del Valle,$^{3}$ F. P. Laussy,$^{3\ast}$ D. Sanvitto$^{1,2\ast}$\\
  \normalsize{$^{1}$Istituto Italiano di Tecnologia, IIT-Lecce, Via Barsanti, 73010 Lecce, Italy}\\
  \normalsize{$^{2}$NNL, Istituto Nanoscienze-CNR, Via Arnesano, 73100 Lecce, Italy}\\
  \normalsize{$^{3}$Condensed Matter Physics Center (IFIMAC), Universidad Aut\'onoma de Madrid, E-28049, Spain.}\\
  \normalsize{$^{4}$Instituto de F\'{i}sica, Universidad de Antioquia, Medell\'{i}n AA 1226, Colombia}\\
  \normalsize{$^{5}$Laboratoire Kastler Brossel, UPMC-Paris 6, ENS et CNRS, 75005 Paris, France}\\
  \normalsize{$^{6}$Universit\'{a} del Salento, Via Arnesano, 73100 Lecce, Italy}\\
  \normalsize{$^\ast$To whom correspondence should be addressed;} \\
  \normalsize{E-mail: lorenzo.dominici@gmail.com; fabrice.laussy@gmail.com}\\
}

\begin{abstract}
  We report the experimental observation and control of space and
  time-resolved light-matter Rabi oscillations in a microcavity.  Our
  setup precision and the system coherence are so high that coherent
  control can be implemented with amplification or switching off of
  the oscillations and even erasing of the polariton density by
  optical pulses.  The data is reproduced by a fundamental quantum
  optical model with excellent accuracy, providing new insights on the
  key components that rule the polariton dynamics.
\end{abstract}

\maketitle

Rabi oscillations~\cite{rabi37a} are the embodiment of quantum
interactions: when a mode~$a$ is excited and is coupled to a second
mode~$b$, the excitation is transferred from~$a$ to~$b$ and when the
symmetric situation is established, the excitation comes back in a
cyclical unitary flow. When this occurs at the single particle level
in a quantum two-level system, it provides the ground for a
qubit~\cite{schumacher95a}, which, if it can be further manipulated,
opens the possibility to perform quantum information
processing~\cite{nielsen_book00a}.  Such an oscillation is of
probability amplitudes and therefore is a strongly quantum mechanical
phenomenon, that involves maximally entangled
states:
\begin{equation}
  \label{eq:lunjul21152706CEST2014}
  \ket{\Psi(t)}=\alpha(t)\ket{1_a,0_b}+\beta(t)\ket{0_a,1_b}\,.
\end{equation}
The same physics also holds, not at the quantum level, but with
coherent states of the fields, a situation known in the literature as
implementing an ``optical atom''~\cite{spreeuw93a} or a ``classical
two-level system''~\cite{faust13a}.  The oscillation is then more
properly qualified as ``normal mode
coupling''~\cite{zhu90a,khitrova06a} as it is now between the fields
themselves:
\begin{equation}
  \label{eq:marjul1094910CEST2014}
  \ket{\psi(t)}=\ket{\alpha(t)}\ket{\beta(t)}\,,
\end{equation}
rather than their probability amplitudes.  The denomination of Rabi
oscillations remains however popular also in this
case~\cite{matthews99b,vasa13a}.  While of limited value for hardcore
implementation of quantum information processing, it is desirable for
fundamental purposes and semi-classical applications to have access to
such classical qubits, or ``cebit''~\cite{spreeuw98a}. In particular,
they can help to explore the origin and mechanism of nonlocality and
parallelization in genuinely quantum systems~\cite{spreeuw01a}, as
well as providing classical counterparts useful for proof-of-principle
demonstration, design and optimization of the actual quantum
version~\cite{dragoman_book04a}. Such classical two-level systems have
been pursued for decades~\cite{spreeuw90a} and recently enjoyed a
boost with the rise of nanomechanical
optics~\cite{faust13a,okamoto13a}. There is another system which
provides an ideal platform to implement both genuinely
quantum~\cite{hennessy07a} and classical versions~\cite{weisbuch92a}
of the two-level system: polaritons~\cite{kavokin_book11a}. A
polariton is by essence a two-level system, arising from strong
light-matter coupling between a cavity photon and a semiconductor
exciton. In planar microcavities, which is the case of interest here,
the system has enjoyed considerable attention for its quantum
properties at the macroscopic level~\cite{carusotto13a}, such as
Bose-Einstein condensation~\cite{kasprzak06a},
superfluidity~\cite{amo09a,amo09b} and a wealth of quantum
hydrodynamics features~\cite{liew08a,amo10a,amo11a}, culminating with
the demonstration of possible devices~\cite{baumberg08a,schneider13a}
and pioneering logical operations~\cite{ballarini13a}. While Rabi
oscillations are at the heart of polariton physics, they are so fast
in a typical microcavity---in the sub-picosecond timerange---that they
are typically glossed over and the macroscopic physics of polaritons
investigated in their coarse graining.  Pioneering attempts to observe
them showed the inherent difficulty and reported hardly two
oscillations with three orders of magnitude loss of contrast each
time~\cite{norris94a}, attributed to the inhomogeneous broadening of
excitons by the theory~\cite{savona96b} which could provide a
qualitative agreement only. Later reports through pump-probe
techniques~\cite{wang95a,marie99a}, in particular in conjunction with
an applied magnetic field~\cite{brunetti06a}, increased their
visibility but remained tightly constrained to their bare
observation. Since polaritons are increasingly addressed at the single
particle level~\cite{boulier14a,arXiv_silva14a}, it becomes capital to
harness their Rabi dynamics.

\begin{figure}[t]
  \centering
  \includegraphics[width=.9\linewidth]{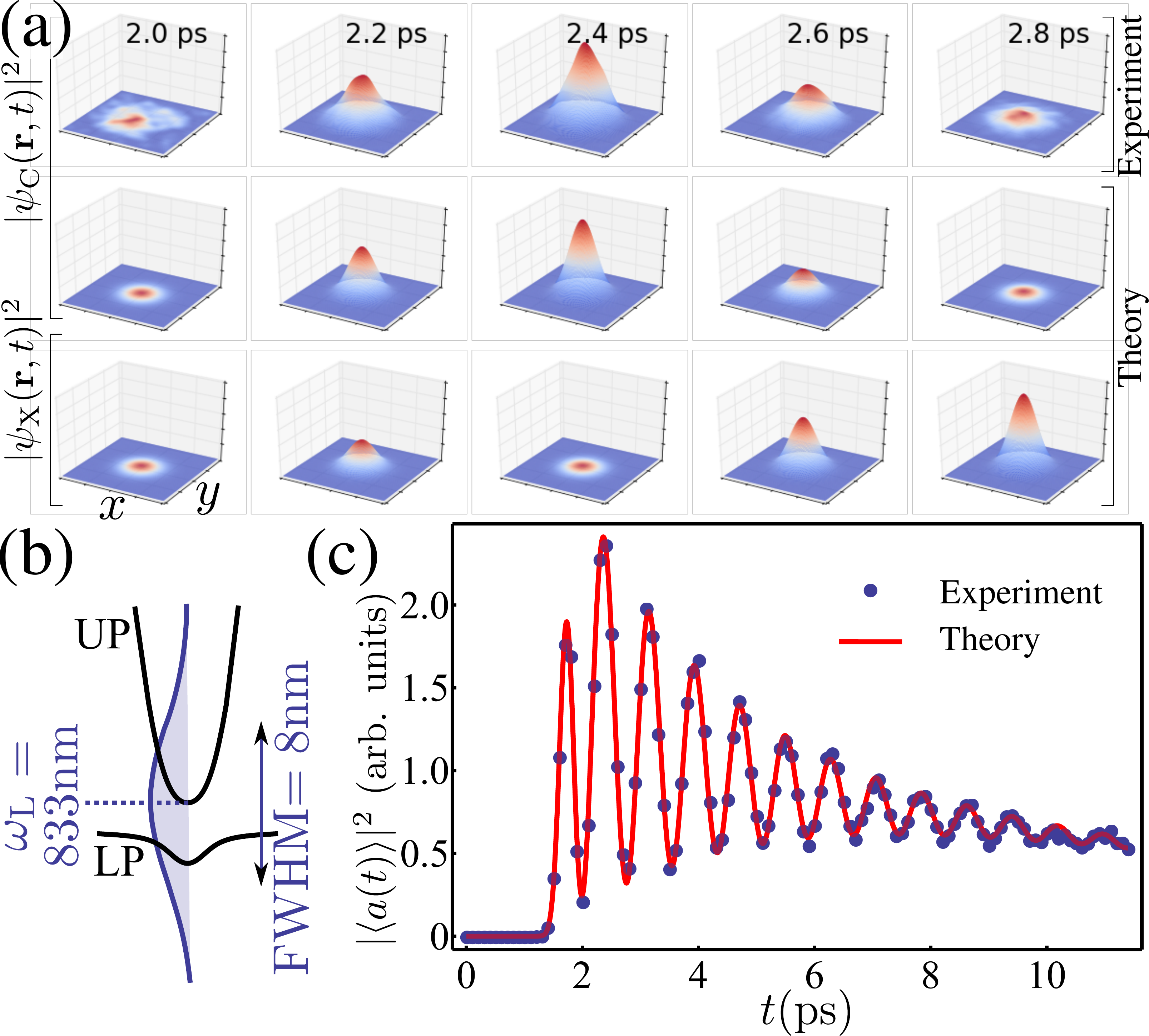}
  \caption{(Color online) (a) Oscillations observed experimentally in
    the cavity field and reproduced theoretically in both the cavity
    and exciton fields. (b) The Lower (LP) \& Upper (UP) Polaritons
    excited by a Gaussian pulse which overlap with the branches
    determines the effective state created in the system.  (c) The
    dynamics can be reduced to that of $|\langle a(t)\rangle|^2$ alone
    and described quantitatively by the theory.}
  \label{fig:1}
\end{figure}

In this Letter, thanks to significant progress in both the quality of
the structures and in the laboratory state of the art, we have been
able to both observe and provide a full control of the microcavity
polariton Rabi dynamics.  This brings microcavities one step further
as platforms to control and engineer various states of light-matter
coupling. We can span from Rabi oscillating configurations to
eigenstate superpositions, and control them by optical pulses that can
amplify or switch states, thereby achieving the same type of coherent
control recently reported in mechanical systems~\cite{faust13a}, but
fully optically and with over nine orders of magnitude gain in speed.
The data offers a perfect quantitative agreement with a fundamental
model of light-matter coupling of two bosonic fields, that allows us
to pin down the underlying dynamics and explain which factors play
which role and to which extent, at the highest level of precision ever
attained in a microcavity, thus making such systems even more suitable
for engineering and applications.

A typical experimental observation is shown in Fig.~\ref{fig:1}(a):
the cavity field oscillates after its excitation by a
$\SI{100}{\femto\second}$ long and $\SI{8}{\nano\meter}$ energy broad
pulse impinging on both branches, as sketched in
Fig.~\ref{fig:1}(b). The basic interpretation is straightforward: by
exciting both branches, the system is prepared as a bare state and,
not being an eigenstate, oscillates between its two components. Since
polaritons are extended objects, the oscillations is between two
fields, localized in a Gaussian of width a few tenths of a $\mu$m
given by the exciting laser. This provides us with the first
observation of the beatings of a ``light-matter drum''. Such a
striking dynamics can be accessed thanks to our ultrafast imaging
technique based on homodyne interferometric detection, described in a
previous work~\cite{arXiv_dominici13a}. This allows us to observe the
subpicosecond Rabi oscillations in the direct emission from the
exciton-polariton fluid through the coherent fraction
$|\psi_a(\mathbf{r},t)|^2$ of the cavity field in both
space~$\mathbf{r}$ and time~$t$.  We used a good quality sample
($\mathrm{Q}\approx14\,000$), providing a cavity lifetime
($\tau_a=\SI{5}{\pico\second}$) and an exciton lifetime
($\tau_b=\SI{1}{\nano\second}$) much longer than the Rabi period,
estimated from the coupling
strength~$g\approx\SI{5.3}{\milli\electronvolt}$ as
$\SI{800}{\femto\second}$. The coupling itself is obtained from the
$\SI{3}{\nano\meter}$ Rabi splitting between the Lower Polariton (LP)
and Upper Polariton (UP) branches.  The power is set to excite
polaritons at a low enough density, in order to maintain their bosonic
properties in the linear regime.

Both the photon-field~$\psi_a$ dynamics of the experiment and the
complementary exciton field $\psi_b$, not accessible experimentally,
can be recovered by the usual polariton field
equations~\cite{ciuti05a} (cf.~Supplementary~\cite{sm}). As expected,
the exciton field forms as the photon field vanishes before it is
revived as the excitations flow back from excitons into photons again.
Limiting to cases with no momentum, although the wavefunction
components have a spread in both real and reciprocal spaces, the
system is linear and there is no dynamics imparted by the spatial
degree of freedom. The dynamics can therefore be reduced to
zero-dimension between two single harmonic modes, and the oscillations
are fully captured through the simpler order parameters $\langle
a(t)\rangle=\int\psi_a(\mathbf{r}, t)d\mathbf{r}$, accessible
experimentally, and $\langle
b(t)\rangle=\int\psi_b(\mathbf{r}, t)d\mathbf{r}$.  This is
shown in Fig.~\ref{fig:1}(c) as points, now for the full duration of
the experiment. Twelve oscillations are clearly resolved until
$t=\SI{12}{\pico\second}$.  Theoretically, the hamiltonian is reduced
to simply $H_0=\hbar\omega_0(\ud{a}a+\ud{b}b)+\hbar
g(\ud{a}b+a\ud{b})$, with coupling strength~$g$ between the photonic
mode~$a$ and the emitter annihilation operator~$b$, both following
Bose algebra and at energy~$\omega_0$ (resonant case), supplemented
with
$H_\Omega=\sum_{c=a,b}P_c(t)e^{i\omega_\mathrm{L}t}e^{i\phi_c}\ud{c}+\mathrm{h.c.}$
%
%
This is the most general case of coherent and resonant excitation,
with coupling to both fields and allowing for a relative phase, which
is necessary to reproduce the data.  Although the excitation is an
optical laser shone directly on the cavity, which is often described
theoretically as a cavity-only coupling
term~\cite{carusotto04a,ciuti05a,cancellieri10a}, it is clear on
physical grounds that such a general form may be required instead,
since the exciton field would still be excited without the cavity so
it is natural that part of the excitation is shared between the latter
and the Quantum Well (QW). While it has little consequence for the
single-pulse excitation, this will be crucial when dealing with
coherent control by a second pulse.  This prepares states of the type
of Eq.~(\ref{eq:marjul1094910CEST2014}), i.e., classical states that
should not be confused with quantum superpositions of the type of
Eq.~(\ref{eq:lunjul21152706CEST2014})~\cite{demirchyan14a}, which
would be extremely difficult to realize and maintain even for small
values of $|\alpha|^2$ and $|\beta|^2$. Regardless of the magnitude of
pumping, the coherent excitation of two linearly coupled oscillators
cancels completely the entanglement of the polariton Fock states to
produce factorizable states, in any basis.  By integrating
Schr\"odinger's equation $i\hbar\partial_t\psi=H\psi$, one easily
finds the closed-form expression for $\alpha(t)$ and $\beta(t)$ under
the dynamics of $H=H_0+H_\Omega$~\cite{sm}. For the case of an initial
state~$\ket{\alpha_0}\ket{\beta_0}$, $\ket{\psi(t)}$ reduces to:
\begin{equation}
  \label{eq:marjul1095135CEST2014}
  \ket{\alpha_0\cos(gt)-i\beta_0\sin(gt)}\ket{-i\alpha_0\sin(gt)+\beta_0\cos(gt)}\,.
\end{equation}
This describes two quantum oscillators, swinging like any other of
their classical counterparts, and that mixes features of the bare
states (which amplitudes oscillate), with those of the polaritons
(with no oscillations of their amplitudes).  From the observation of
the oscillation alone, it is therefore difficult to capture the true
dynamics at play. This is where a theoretical model is needed to shed
light on the hidden features~\cite{laussy09a}. As we are going to
show, the contrast of the oscillation is not due to decoherence
between the UP and LP, but to a combination of the short lifetime of
the UP and of the effective state realized by the pulsed excitation.

\begin{figure}[t]
  \centering
  \includegraphics[width=.9\linewidth]{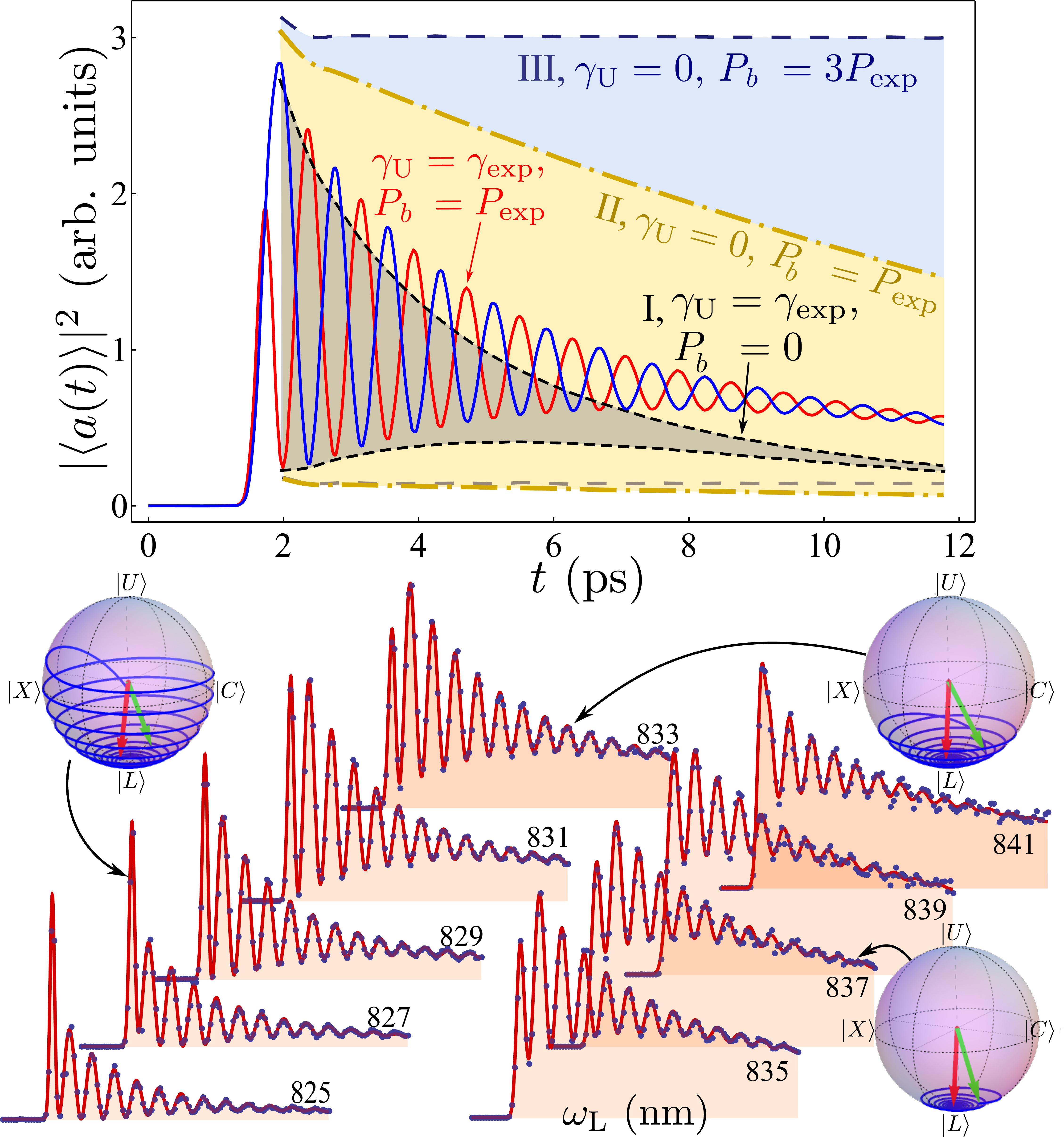}
  \caption{(Color online) Various states created in the system by
    varying the pulse energy and their evolution on the Bloch sphere,
    showing the systematic relaxation towards the LP. In inset, the
    full exciton-photon dynamics reconstructed theoretically for the
    case $\omega_\mathrm{L}=\SI{829}{\nano\meter}$ and variations
    (showing the envelope of the oscillations only) when removing the
    effect of the exciton reservoir (I, black dashed), removing the
    effect of polariton dephasing (II, dotted-dashed yellow) and
    removing both (III, long-dashed blue). $P_\mathrm{exp}$ and
    $\gamma_\mathrm{exp}$ are the fitted values for the experiment.}
  \label{fig:2}
\end{figure}

While the core of the physics is contained in the wavefunction
$\ket{\psi}$ of the coupled oscillators under the dynamics of $H$, we
have to take into account dephasing and decay to describe any
experiment with some degree of accuracy. These are mainly due to the
bare state lifetimes (with decay rates $\gamma_{a,b}$ for the
photon/exciton, respectively), which are also present in most
light-matter coupled systems. In QW microcavities, additional sources
of dephasing are present for the UP, which is well known to be much
less visible than its LP counterpart~\cite{baumberg98a,skolnick00a}. A
contribution from the exciton reservoir has also been suggested in
several works, even under coherent
excitation~\cite{vishnevsky12a,wouters13a}. However, no direct
measurement of its contribution, nor its true nature (coherent or
incoherent) has been clearly reported until now.  We can address this
issue by including an UP dephasing rate $\gamma_\mathrm{U}$ and an
incoherent excitonic pumping rate $P_b$. Indeed, both terms are
required to reproduce the data at the level of accuracy we report.
Such terms turn the pure state wavefunction into a density matrix
$\rho$ ruled by a master equation. The theory is standard and is given
in the Supplementary material~\cite{sm}. In this case, the complex
amplitudes of the oscillators can also be derived in closed-form
expressions.  The experimental modulus square of the cavity amplitude
can then be fitted by the model and other observables reconstructed
from the theory. The fit provides an essentially perfect agreement
with the data, as seen in the figures.

By shifting the laser energy to weight more on one branch than the
other, as done for the series displayed in Fig.~\ref{fig:2}(a),
different states can be prepared, that are all equally well accounted
for by the theory for the same system parameters.  Note also that both
the dynamics of the pulse as well as the subsequent free oscillations
is described within the same model.  From the theory fitting the
experimental $|\langle a(t)\rangle|^2$, we gain access to the entire
dynamics of oscillations, also of the exciton field $|\langle
b(t)\rangle|^2$, but even further, of the phases $\langle a(t)\rangle$
and $\langle b(t)\rangle$ and the total excitations
$\langle(\ud{a}a)(t)\rangle$ and $\langle(\ud{b}b)(t)\rangle$ and, in
fact, of the full state as a whole through the density matrix
$\rho$. This allows us to reconstruct the full dynamics, as done in
Fig.~\ref{fig:2} for the joint exciton-photon oscillations of the
experiment (case of \SI{827}{\nano\meter} excitation), and see the
effect of the various factors involved. For instance, the impact of
the reservoir is seen in the case~I (in dashed black line, from now on
plotting only the envelope of the Rabi oscillations for clarity) where
it has been set to zero. Its effect is small but is needed to
reproduce the data quantitatively. The main detrimental actor is the
UP dephasing rate~$\gamma_\mathrm{U}$, which, if set to zero,
considerably opens the envelope of oscillations (case II, blue
dashed-dotted line). Interestingly, the incoherent reservoir extends
the lifetime of the oscillations as shown in case II and even more so
in case~III (long dashed line) where an higher pumping rate than that
of the experiment brings the oscillations well into the nanosecond
timescale, as proposed in Ref.~\cite{demirchyan14a}, although, as
already noted, this is for normal mode coupling oscillations that
cannot be used to engineer a qubit.

The coherent amplitudes of any two-level system can be mapped on the
Bloch sphere as $\langle a\rangle/\sqrt{|\langle
  a\rangle|^2+|\langle b\rangle|^2}=\cos(\theta/2)$ and $\langle
b\rangle/\sqrt{|\langle a\rangle|^2+|\langle
  b\rangle|^2}=\sin(\theta/2)\exp(i\phi)$ with $\theta$ and $\phi$ the
azimuthal and radial angles of polar coordinates, respectively.  Such
trajectories from our experiment are shown for three cases in
Fig.~\ref{fig:2}, corresponding to predominant UP excitation, equal
weight of the branches and predominant LP excitation. It is clearly
seen in the first case how the pulse swings the coupled oscillators
towards the upper state and, in all cases, how the system quickly
reaches the LP. This is the clearest observation to date of one of the
most important assumptions of microcavity polariton physics: the UP is
unstable and the system relaxes towards the lower branch, even though
it retains strong-coupling. In the model, this UP dephasing rate
$\gamma_\mathrm{U}$, could be either an escape rate (like a lifetime
due to, e.g., scattering to high-$k$ exciton states), a pure dephasing
rate, or a combination of boths, as only their sum enters in the
equation of the coherent fraction.  The result also shows that
although the impinging laser is very wide in energy, it is possible to
prepare the polariton condensate in a largely tunable range, from
almost entirely upper-polaritonic (at least for short times) to almost
entirely lower-polaritonic (also the state at long-times), passing by
purely photonic and/or excitonic, these two states constantly
oscillating between each other.

\begin{figure}[t]
  \centering
  \includegraphics[width=\linewidth]{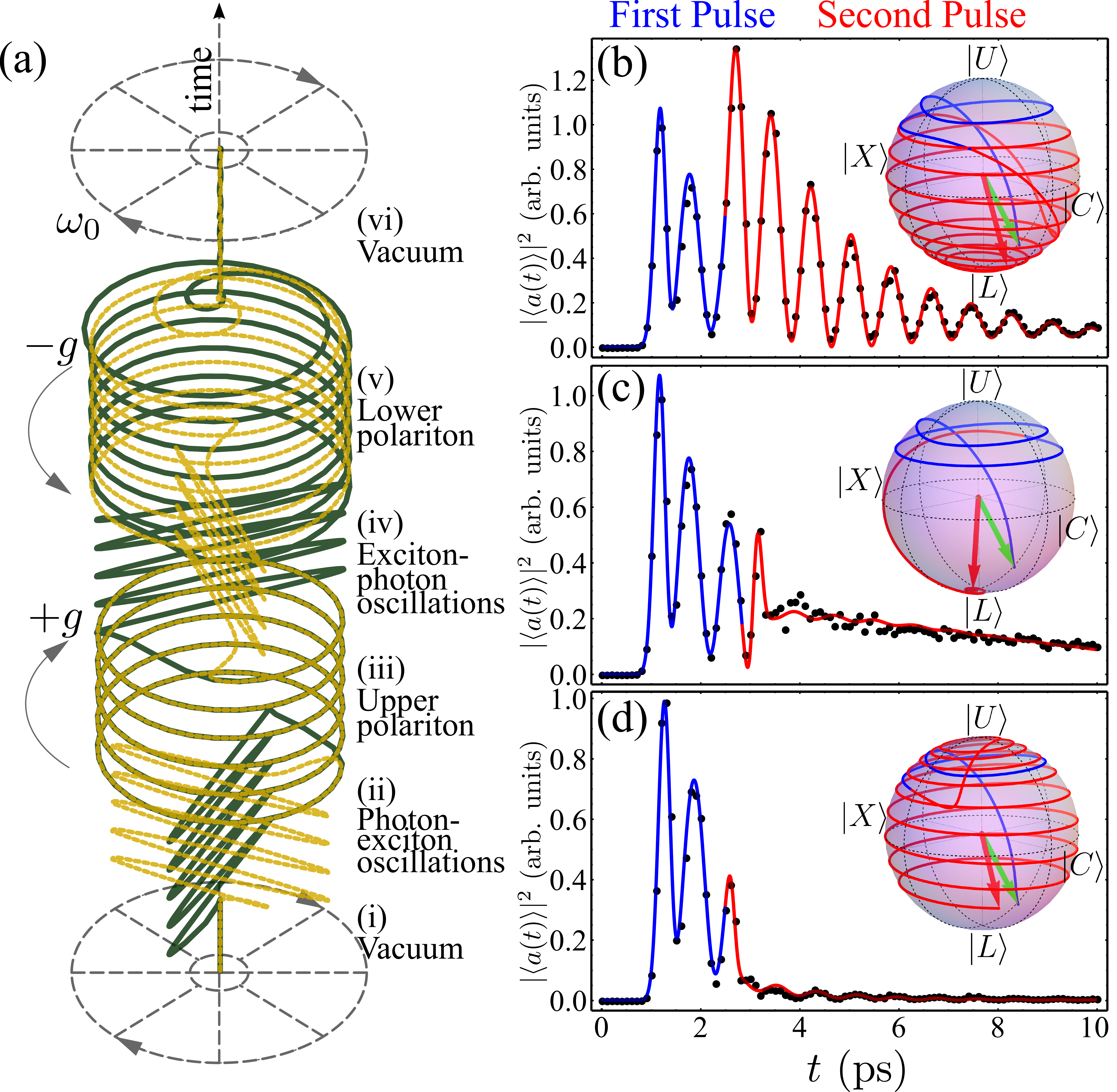}
  \caption{(Color online) (a) Succession of various dynamical
    evolutions in time of the cavity (dashed yellow) and the exciton
    (green) complex amplitudes (here with fixed modulus) that can be
    passed from one to the other with an appropriate pulse excitation.
    Normal mode-coupling features a $\pi/2$ dephasing in both time and
    optical phase and oscillate radially while polaritons oscillate
    circularly. Furthermore, they oscillate jointly and against
    (resp.~$\pi$ out-of-phase with each other and with) the rotating
    frame in the case of a LP (resp.~UP).  (b--e) Experimental
    realization (points) and theoretical fit (solid curve) of three
    two-pulses excitation, showing (b) amplification, (c) transition
    from an exciton-photon Rabi oscillation to a LP and
    (d) field annihilation.}
  \label{fig:3}
\end{figure}

With such an accurate command of the system, we are able to time
precisely the arrival of a second pulse and perform a comprehensive
coherent control on the coupled dynamics.  For a coupling of the laser
to the cavity only, this would be achieved for most operations by
sending the control pulse when the cavity field is empty and the
condensate fully excitonic. Injecting a second fully photonic pulse in
optical (resp.~anti-optical) phase with the exciton, for instance,
creates an UP (resp.~LP).  It is convenient to represent such a
dynamics with the joint photon and exciton fields' complex phases, as
shown in Fig.~\ref{fig:3}(a) for a sequence of basic operations
through pulsed excitation that bring the system from (i) the vacuum
and (vi) back passing by a condensate of (ii) photons, (iii) UPs, (iv)
excitons and (v) LPs. The photon and exciton condensates are defined
as such right after the pulse only since, not being eigenstates, they
enter the oscillating regime. In the rotating frame of the bare modes
at frequency~$\omega_0$, the light-matter dynamics is a simple
oscillation along the radius with a combined offset of $\pi/2$ both in
time and optical phase: the cavity oscillates horizontally while the
exciton oscillates vertically and when one reaches its maximum, the
other is at the origin. In contrast, the LP and UP condensates do not
oscillate radially but circularly, since they are free modes that
subtract and add, respectively, their free energy to that of the
rotating frame. An animation of this dynamics is given in the
Supplementary Material~\cite{sm}.

In the actual experiment, where the laser couples to both fields, one
merely needs to correct for the corresponding proportions but the
concept is otherwise the same. A first pulse triggers the Rabi
oscillations, since our pulse is broad in energy and always initiate a
dominant photon or exciton fraction. However, with a second pulse,
although still broad, we can refine the state by providing the
complementary of the sought target. Figure~\ref{fig:3}(b) shows a
simple case of Rabi amplification, where the same cycle is restarted
by the pulse. Figure~\ref{fig:3}(c) shows the case where a bare state
is transformed into a LP, therefore switching off the
oscillations.  While it has not been practical with our setup to send
more than two pulses, there is no fundamental difficulty in doing so
and in principle one can implement all the wished steps to prepare any
given state right after the pulses. Another case of interest is
complete field annihilation, by sending a pulse out-of-phase both in
amplitude and in optical phase. This produces, by destructive
interferences, the vacuum, as shown in Fig.~\ref{fig:3}(d). All these
cases demonstrate the possibility to do coherent control of the
strong light-matter coupling dynamics. Here too, the theory still
provides an essentially perfect agreement to the data. Similar
prospects at the single-particle level would perform genuine quantum
information processing, but this lies beyond the scope of this work.

In conclusion, we showed the tremendous control that can be obtained
on the light-matter coupling in microcavities, for which we reported
the first imaging of its spatio-temporal evolution. This allowed us to
spell out with a precision never achieved before for polaritons both
the excitation scheme and the various components involved in the
dynamics (dephasing, reservoirs, etc.) We demonstrated the
reservoir-induced lifetime enhancement recently
proposed~\cite{demirchyan14a} and performed coherent control on the
polariton state. Such results are a milestone to turn these systems
into devices, with future prospects such as optical gates or their
single-particle counterpart now clearly in sight. Immediate extensions
suggested by this work are---beyond getting to the single-particle
limit---to couple to the spatial degree of freedom with packets
imparted with momentum or diffusing, and involve nonlinearities at
higher pumpings.

Funding by the ERC POLAFLOW and the IEF SQUIRREL is acknowledged.


\pagebreak

\begin{widetext}

\begin{center}
\textbf{\large Ultrafast control of Rabi oscillations in a polariton condensate : \\ 
Supplementary Material.}
\end{center}

\begin{center}
{L. Dominici,$^{1,2\ast}$ D. Colas,$^{3}$ S. Donati,$^{1,2}$ J.~P. Restrepo Cuartas,$^{3}$ M. De Giorgi,$^{1,2}$  D. Ballarini,$^{1,2}$ G. Guirales,$^{4}$\\J. C. L\'opez Carre\~no,$^{3}$ A. Bramati,$^{5}$ G. Gigli,$^{1,2,6}$ E. del Valle,$^{3}$ F. P. Laussy,$^{3\ast}$ D. Sanvitto$^{1,2\ast}$\\
  \normalsize{$^{1}$Istituto Italiano di Tecnologia, IIT-Lecce, Via Barsanti, 73010 Lecce, Italy}\\
  \normalsize{$^{2}$NNL, Istituto Nanoscienze-CNR, Via Arnesano, 73100 Lecce, Italy}\\
  \normalsize{$^{3}$Condensed Matter Physics Center (IFIMAC), Universidad Aut\'onoma de Madrid, E-28049, Spain.}\\
  \normalsize{$^{4}$Instituto de F\'{i}sica, Universidad de Antioquia, Medell\'{i}n AA 1226, Colombia}\\
  \normalsize{$^{5}$Laboratoire Kastler Brossel, UPMC-Paris 6, ENS et CNRS, 75005 Paris, France}\\
  \normalsize{$^{6}$Universit\'{a} del Salento, Via Arnesano, 73100 Lecce, Italy}\\
  \normalsize{$^\ast$To whom correspondence should be addressed;} \\
  \normalsize{E-mail: lorenzo.dominici@gmail.com; fabrice.laussy@gmail.com}\\
}
\end{center}

\begin{small}
  In this supplementary material, we provide details on the
  theoretical model, the nature of the state realized in the
  experiment and the problem of its visualization, the effect of pure
  dephasing, some limiting analytical solutions and additional fitted
  data.
\end{small}

\end{widetext}

\setcounter{equation}{0}
\setcounter{figure}{0}
\setcounter{table}{0}
\setcounter{page}{1}
\makeatletter
\renewcommand{\theequation}{S\arabic{equation}}
\renewcommand{\thefigure}{S\arabic{figure}}
\renewcommand{\bibnumfmt}[1]{[S#1]}
\renewcommand{\citenumfont}[1]{S#1}

\section{Rabi oscillations between two extended spatial fields}

Polaritons in the linear regime implement the physics of two-coupled
Schr\"odinger equations~\cite{S_laussy12a}:
\begin{subequations}
  \label{eq:maroct15123657CEST2013}
  \begin{align}
    i\hbar\partial_t\psi_a(\mathbf{r},t)&=-\frac{\hbar^2\nabla^2\psi_a(\mathbf{r},t)}{2m_a}+\hbar g\psi_b(\mathbf{r},t)\,,\\3
    i\hbar\partial_t\psi_b(\mathbf{r},t)&=-\frac{\hbar^2\nabla^2\psi_b(\mathbf{r},t)}{2m_b}+\hbar g\psi_b(\mathbf{r},t)\,.
  \end{align}
\end{subequations}
Each coupled field (cavity photon and exciton) serves as a potential
for the other. Here $m_{a,b}$ are the cavity photon
and exciton masses, respectively, $g$ their coupling strength, and we
are ignoring dissipation for a while.  Interestingly, such a simple
equation has to the best of our knowledge no known analytical
solution. While the coupling can give rise to a complicated
dynamics~\cite{S_unpub_colas14a}, when the diffusion is small and in the
linear regime, the phenomenology is simply that of two fields beating
at the Rabi frequency. In our experiments, we have provided the first
observation of the dynamics of such a ``drum of light-and-matter'',
cf.~Fig.~1(a) of the main text, and our ability to control some of its
modes of motion. The Supplementary Video
\texttt{I-SpaceTimeRabiOscillation.avi}~\cite{S_movieI} shows the
dynamics along the diameter (along a 1D cut of the 2D spatial
dynamics) and time for the pulsed excitation at \SI{831}{\nano\meter}.

In our configuration, where the space dynamics plays no substantial
role, it can be averaged over to retain only the key variables that
capture the dynamics, namely:
\begin{equation}
  \langle a(t)\rangle=\int\psi_a(\mathbf{r},t)\,d\mathbf{r}\quad
  \mathrm{and}\quad
  \langle b(t)\rangle=\int\psi_b(\mathbf{r},t)\,d\mathbf{r}\,,
\end{equation}
in which case the dynamics becomes that of two coupled quantum
harmonic oscillators, with commutation relation $[c,\ud{c}]=1$ for
both~$c=a,b$, and Hamiltonian (at resonance):
\begin{equation}
  \label{eq:miejun25183432CEST2014}
  H=\hbar\omega_a(\ud{a}a+\ud{b}b)+\hbar g\left(\ud{a}b+a\ud{b}\right)\,.
\end{equation}

\section{Hamiltonian dynamics of two linearly coupled oscillators}

While the dynamics of two linearly coupled oscillators is extremely
simple from a classical perspective, it can become tricky when
shifting to the quantum point of view. In addition to the various
patterns of oscillations, just as in the classical case, one has to
consider the plethora of quantum states that can set them in motion
(or not, in cases of eigenstate superpositions). Already for pure
states, the wavefunction needs an infinite set of complex
coefficients~$\alpha_{nm}(t)$ to be fully specified, say in the basis
of Fock states~$\ket{nm}$ with $n$ quanta of excitations in
oscillator~$a$ and~$m$ in oscillator~$b$, since:
\begin{equation}
  \label{eq:juejun5193254CEST2014}
  \ket{\psi(t)}=\sum_{n,m=0}^\infty\alpha_{nm}(t)\ket{nm}\,.
\end{equation}

While as much information is necessary to describe genuinely quantum
states~\cite{S_unpub_carreno14a}, most of it is redundant for those that
have a classical counterpart. For instance, coherent states can be
reduced to two complex numbers only:
\begin{equation}
  \label{eq:miéjun25182235CEST2014}
  \ket{\psi(t)}=\ket{\alpha(t)}\ket{\beta(t)}\,.
\end{equation}

This is essentially the case of our experiment since the system is
excited resonantly by a laser pulse, that creates a coherent state,
and it is found a-posteriori by the analysis that dephasing has a
small effect given the large populations involved and the short
duration of the experiment, resulting in states that remain
essentially fully coherent throughout
(cf.~Section~\ref{sec:domjun29114211CEST2014}).

Adding the excitation scheme as:
\begin{equation}
  \label{eq:miejun4104129CEST2014}
  H_\Omega=\sum_{c=a,b}P_c(t)e^{i\omega_\mathrm{L}t}e^{i\phi_c}\ud{c}+\mathrm{h.c.}
\end{equation}
one can find by straightforward algebra the wavefunction at all later
times~$t$ under the dynamics of Eq.~(\ref{eq:miejun25183432CEST2014}),
starting from the initial condition:
\begin{equation}
  \label{eq:viejun27093049CEST2014}
  \ket{\psi(t=0)}=\ket{\alpha_0}\ket{\beta_0}\,,
\end{equation}
for any~$\alpha_0$, $\beta_0\in\mathbf{C}$ (possibly zero), to be
Eq.~(\ref{eq:miéjun25182235CEST2014}) with $\alpha(t)$ and $\beta(t)$
given by:
\begin{align}
  \label{eq:miejun25185414CEST2014}
  \gamma(t)=&\gamma_0\cos(gt)-i\bar\gamma_0\sin(gt) \\ \nonumber
  -&G^+\exp(-igt)\left(\mathrm{erfi}\left(\frac{it_1+\tau^+}{\sqrt{2}\sigma}\right)-\mathrm{erfi}\left(\frac{-i(t-t_1)+\tau^+}{\sqrt{2}\sigma}\right)\right)\\
  \nonumber
  \mp&G^-\exp(igt)\left(\mathrm{erfi}\left(\frac{it_1+\tau^-}{\sqrt{2}\sigma}\right)-\mathrm{erfi}\left(\frac{-i (t-t_1)+\tau^-}{\sqrt{2}\sigma}\right)\right)  \end{align}
with $\gamma$ a notation for either~$\alpha$ or $\beta$ with the
convention that $\bar\alpha=\beta$ and $\bar\beta=\alpha$, where we
have also defined:
\begin{subequations}
  \label{eq:domjun29140316CEST2014}
  \begin{align}
    G^\pm=&\frac{1}{4}\exp\left(-\frac{\tau^\pm(2it_1+\tau^\pm)}{2\sigma^2}\right)(P_a\pm P_b\exp(i\phi))\,,\\
    \tau^\pm=&\sigma^2(\mp g+\omega_l)\,.
  \end{align}
\end{subequations}
We have assumed the case of a Gaussian pulse
$P_c(t)=P_ce^{i\phi_c}\exp(-(t-t_1)^2/\sigma^2)/(\sigma\sqrt{2\pi})$
with $\phi=\phi_b$ and~$\phi_a=0$ with no loss of generality.  This is
the key physics at play in our experiment, and we now consider it in
more details.

A first important clarification in the light of a recently published
proposal~\cite{S_demirchyan14a} is that the state that is created
following such an excitation is a factorizable product of coherent
states of the type~(\ref{eq:miéjun25182235CEST2014}), and not a
quantum superposition of ``\emph{macroscopically} occupied orthogonal
states''. It is true that any Fock state of polaritons is in a such
quantum superposition, starting with the building blocks of
one-particle states:
\begin{subequations}
  \begin{align}
    \label{eq:juemay8192800CEST2014}
    \kket{1,0}&=\cos\theta\ket{0,1}-\sin\theta\ket{1,0}\,,\\
    \kket{0,1}&=\sin\theta\ket{0,1}+\cos\theta\ket{1,0}\,,
  \end{align}
\end{subequations}
writing $\ket{n,m}$ as a shortcut for~$\ket{n}\ket{m}$ the pure state
with $n$ photons and $m$ excitons, and $\kket{n,m}$ the pure state
with $n$ lower polaritons and $m$ upper polaritons. Regardless of the
quantum nature of these one-particle states, however, coherent
superpositions of them wash out the entanglement and result in
coherent states, in any basis. For instance, a coherent state of lower
polaritons:
\begin{equation}
  \label{eq:juemay8200321CEST2014}
  \kket{\alpha,0}=e^{-|\alpha|^2/2}\sum_{n=0}^\infty\frac{\alpha^n}{\sqrt{n!}}\kket{n,0}\,,
\end{equation}
is a product of coherent states of photons and excitons:
\begin{align}
  \kket{\alpha,0}&=e^{-\frac{|\alpha|^2}{2}}\sum_{n=0}^\infty\frac{\alpha^n}{\sqrt{n!}}\sum_{k=0}^n\sqrt{{n\choose k}}\cos^k\theta\sin^{n-k}\theta\ket{k,n-k}\,,\nonumber\\
  &=\ket{\alpha\cos\theta,\alpha\sin\theta}\label{eq:viemay9004427CEST2014}\,.
\end{align}
It is therefore useless for actual quantum information
processing, describing an essentially classical phenomenon, although
it can have some value to simulate in a controlled classical
environment a model qubit~\cite{S_spreeuw93a}.

If one leaves aside the pulse preparation in
Eq.~(\ref{eq:miejun25185414CEST2014}) to consider directly the
coherent state $\ket{\alpha_0}\ket{\beta_0}$ as the initial condition, the
solution reads simply:
\begin{multline}
  \label{eq:miejun25184605CEST2014}
  \ket{\psi(t)}=\\\ket{\alpha_0\cos(gt)-i\beta_0\sin(gt)}\ket{-i\alpha_0\sin(gt)+\beta_0\cos(gt)}\,.
\end{multline}

Two extreme cases are of interest: on the one hand, $\alpha_0=\mp\beta_0$,
in which case the dynamics becomes
\begin{equation}
  \label{eq:miéjun25200320CEST2014}
  \ket{\psi(t)}=\ket{\alpha_0 e^{\pm igt}}\ket{\mp\alpha_0 e^{\pm igt}}\,,
\end{equation}
corresponding to the freely propagating lower\break
$\kket{-\sqrt{2}\alpha e^{igt},0}$ and upper $\kket{0,\sqrt{2}\alpha
  e^{-igt}}$ polariton condensates, respectively. On the other hand,
for $\alpha$ nonzero and~$\beta=0$ (or vice versa):
\begin{equation}
  \label{eq:miéjun25200955CEST2014}
  \ket{\psi(t)}=\ket{\alpha_0\cos(gt)}\ket{-i\alpha_0\sin(gt)}\,,
\end{equation}
corresponding to Rabi oscillations (here we remind that we use such a
qualification as a convenience and in accord with common usage to
describe what really is normal-mode coupling oscillations) between the
two fields. Polariton states of the
type~(\ref{eq:miejun25184605CEST2014}) oscillate in a circle in the
counter-rotating (resp.~rotating) frame for the lower (resp.~upper)
polariton case (in the non-rotating frame they oscillate faster,
resp.~slower, by a factor~$g$ over the optical frequency~$\omega_a$,
with $g\ll\omega_a$), the two fields being in phase (resp.~$\pi$ out
of phase). In contrast, states of the
type~(\ref{eq:miéjun25200955CEST2014}) oscillate radially and
perpendicularly, the two fields being $\pi/2$ out-of-phase.  The
general case combines these two types of motions and results in an
elliptical oscillation in phase-space, therefore with a reduced
contrast of the oscillations. This is is precisely what happens in our
experiment: the oscillations are not full-amplitude not because of
decoherence that dephase the coupling, as in the case of genuine Rabi
oscillations, but because the state has a strong polaritonic
component. We come back to the problem of visualizing this dynamics in
Section~\ref{sec:domjun29135759CEST2014}, although with the solution
Eqs.~(\ref{eq:miéjun25182235CEST2014}---\ref{eq:domjun29140316CEST2014}),
we have provided a complete solution to the problem. Before
illustrating particular cases of interest, in next Section, we
generalize them to include the effect of dissipation.

\section{Dissipative dynamics of two linearly coupled oscillators}

The Hamiltonian is not enough to quantitatively describe a polariton
experiment, that has various sources of decay. This turns, to begin
with, the wavefunction into a density matrix ruled by von Neumann
equation:
\begin{equation}\label{VonNeumann}
  \dot{\rho} = \frac{i}{\hbar} \left [\rho, H \right ] + \mathcal{L}\rho,
\end{equation}
with $\mathcal{L}\rho$ the Lindblad super-operator which for a generic
operator~$c$ reads:
\begin{equation}
  \label{Lindblad}
  \mathcal{L}_c \rho = 2 c\rho\ud{c} - \ud{c}c\rho - \rho\ud{c}c\,,
\end{equation}
and in our model takes the form~\cite{S_delvalle_book10a}:
\begin{equation}
  \mathcal{L}\rho = \frac{\gamma_a}{2}\mathcal{L}_a\rho + \frac{\gamma_b}{2}\mathcal{L}_b\rho + \frac{\gamma_\mathrm{U}^\mathrm{R}}{2}\mathcal{L}_{u}\rho+\frac{\gamma_\mathrm{U}^\phi}{2}\mathcal{L}_{\ud{u}u}\rho+\frac{P_be^{-\gamma_{P_b}t}}{2}\mathcal{L}_{\ud{b}}\rho\,,
\end{equation}
where $\gamma_{a(b)}$ are the modes $a$ and $b$ radiative lifetimes,
respectively, $\gamma_\mathrm{U}^R$ and $\gamma_\mathrm{U}^\phi$ the
radiative and pure dephasing rates for the upper polariton
$u=(a+b)/\sqrt{2}$ and $P_be^{-\gamma_{P_b}}t$ the incoherent pumping
rate from an exciton reservoir with lifetime $\gamma_{P_b}$ (we will
also introduce late $l=(a-b)/\sqrt{2}$). Such a description is at a
high abstract level but a full microscopic description would be so
complex that it would be impossible without any phenomenological
simplification of this type given the numerical state of the art and
the basic physics that nevertheless eventually takes place.

Here again, most of the information encoded in the density matrix is
not-needed and the dynamics can be reduced to much fewer variables.
To fit the data, since only $|\langle a(t)\rangle|^2$ is accessible
experimentally, it is enough to calculate the dynamics of $\langle
a(t)\rangle$, which only requires that of $\langle b(t)\rangle$. For
the general case of two-pulses excitations, it is therefore enough to
solve numerically their coupled set of equations:
\begin{widetext}
  \begin{subequations}
    \label{eq:viejun27093513CEST2014}
    \begin{align}
      \partial_t\langle a(t)\rangle&=-i( \frac{P_{a}^{(1)}\, e^{i \phi_1} \,e^{-\frac{1}{2}(\frac{t-t_1 }{\sigma_1})^2}}{\sqrt{2 \pi} \sigma_1} +  \frac{P_{a}^{(2)}\, e^{i \phi_2} \,e^{-\frac{1}{2}(\frac{t-t_2 }{\sigma_2})^2}}{\sqrt{2 \pi} \sigma_2})+\frac{1}{4}(-2\gamma_a-\gamma_{\mathrm{U}}+4 i\, \omega_\mathrm{L})\langle a(t)\rangle - \frac{1}{4}(4 i \,g+\gamma_{\mathrm{U}})\langle b(t)\rangle,\\
      \partial_t\langle b(t)\rangle&=-i( \frac{P_{b}^{(1)} e^{-\frac{1}{2}(\frac{t-t_1 }{\sigma_1})^2}}{\sqrt{2 \pi} \sigma_1}+\frac{P_{b}^{(2)} \,e^{-\frac{1}{2}(\frac{t-t_2 }{\sigma_2})^2}}{\sqrt{2 \pi} \sigma_2} ) -\frac{1}{4}(4 i \,g+\gamma_{\mathrm{U}})\langle a(t)\rangle +\frac{1}{4}(2P_b e^{-\gamma_{P_b}\,t}H(t-t_1)-2\gamma_b-\gamma_{\mathrm{U}}+4i\omega_\mathrm{L})\langle b(t)\rangle,
    \end{align}
  \end{subequations}
  with the vacuum as initial condition, that is, $\langle
  a(0)\rangle=\langle b(0)\rangle=0$, and $P_c^{(i)}$ the first,
  $i=1$, and second, $i=2$, pulse coupling to the cavity, $c=a$, and
  exciton field, $c=b$.  Equations~(\ref{eq:viejun27093513CEST2014})
  are those that fit all the data reported in this work. By setting to
  zero the second pulse, $P_a^{(2)}=P_b^{(2)}=0$, we describe the case of
  one pulse excitation. In the case where the exciton reservoir has
  infinite lifetime, $\gamma_{P_b}=0$, this can be integrated
  analytically although the expression is too heavy to be written
  there. Making the further simplification to dispense from the pulse
  dynamics and considering the initial state injected instead,
  Eq.~(\ref{eq:viejun27093049CEST2014}), we can reduce the dynamics to
  a simple form that captures most of the phenomenology:
\begin{subequations}
    \label{eq:juejun5211819CEST2014}
    \begin{align}
      \langle a(t)\rangle &= \begin{aligned}[t]
        &\Bigg[a_0\cosh(\frac{1}{4} R t) -\left(\frac{b_0 G + a_0
            \Gamma}{R}\right)\sinh(\frac{1}{4} R t)\Bigg]\exp(-\frac{1}{4}\gamma t)\,,
      \end{aligned}\\
      \langle b(t)\rangle &=\begin{aligned}[t]
        &\Bigg[b_0\cosh(\frac{1}{4} R t) +\left(\frac{-a_0 G + b_0
            \Gamma}{R}\right)\sinh(\frac{1}{4} R t)\Bigg]\exp(-\frac{1}{4}\gamma t)\,,
      \end{aligned}
    \end{align}
  \end{subequations}
\end{widetext}
where we have introduced:
\begin{subequations}
  \label{eq:juejul10095249CEST2014}
  \begin{align}
    \gamma=& \gamma_a+\gamma_b+\gamma_\mathrm{U}-P_b\,,\\
    \Gamma=& P_b -\gamma_b+\gamma_a\,,\\
    G =& i4g +\gamma_\mathrm{U}\,,\\
    R =& \sqrt{G^2+\Gamma^2}\,,
  \end{align}
\end{subequations}
and, since only the sum of radiative
decay~$\gamma_\mathrm{U}^\mathrm{R}$ and pure
dephasing~$\gamma_\mathrm{U}^\mathrm{\phi}$ of the upper polariton
plays a role in the coherent dynamics, the total upper polariton
dephasing rate:
\begin{equation}
  \label{eq:lunjun30100400CEST2014}
  \gamma_\mathrm{U}=\gamma_\mathrm{U}^\mathrm{R}+\gamma_\mathrm{U}^\phi\,.
\end{equation}

The solution
Eqs.~(\ref{eq:juejun5211819CEST2014}--\ref{eq:juejul10095249CEST2014})
relates to that of
Eqs.~(\ref{eq:miéjun25182235CEST2014}--\ref{eq:domjun29140316CEST2014})
in that it includes dissipation (both radiative lifetime and pure
dephasing of the upper polariton) as well as the incoherent pumping
from the reservoir (assumed constant) but neglects the coherent
excitation (the two pulses), considering an initial state instead. It
also provides the dynamics of $\langle a(t)\rangle$ and $\langle
b(t)\rangle$ only but makes no restriction on the quantum state, that
can be any density matrix, including states with no classical
counterparts, whereas
Eqs.~(\ref{eq:miéjun25182235CEST2014}--\ref{eq:domjun29140316CEST2014})
provides the full quantum state but in a case where it is at all times
in the form Eq.~(\ref{eq:miéjun25182235CEST2014}). Therefore, they
provide two closed-form solutions in two limiting cases. Of course
they agree in their region of overlap.

\begin{figure*}
  \centering
  \includegraphics[width=.85\linewidth]{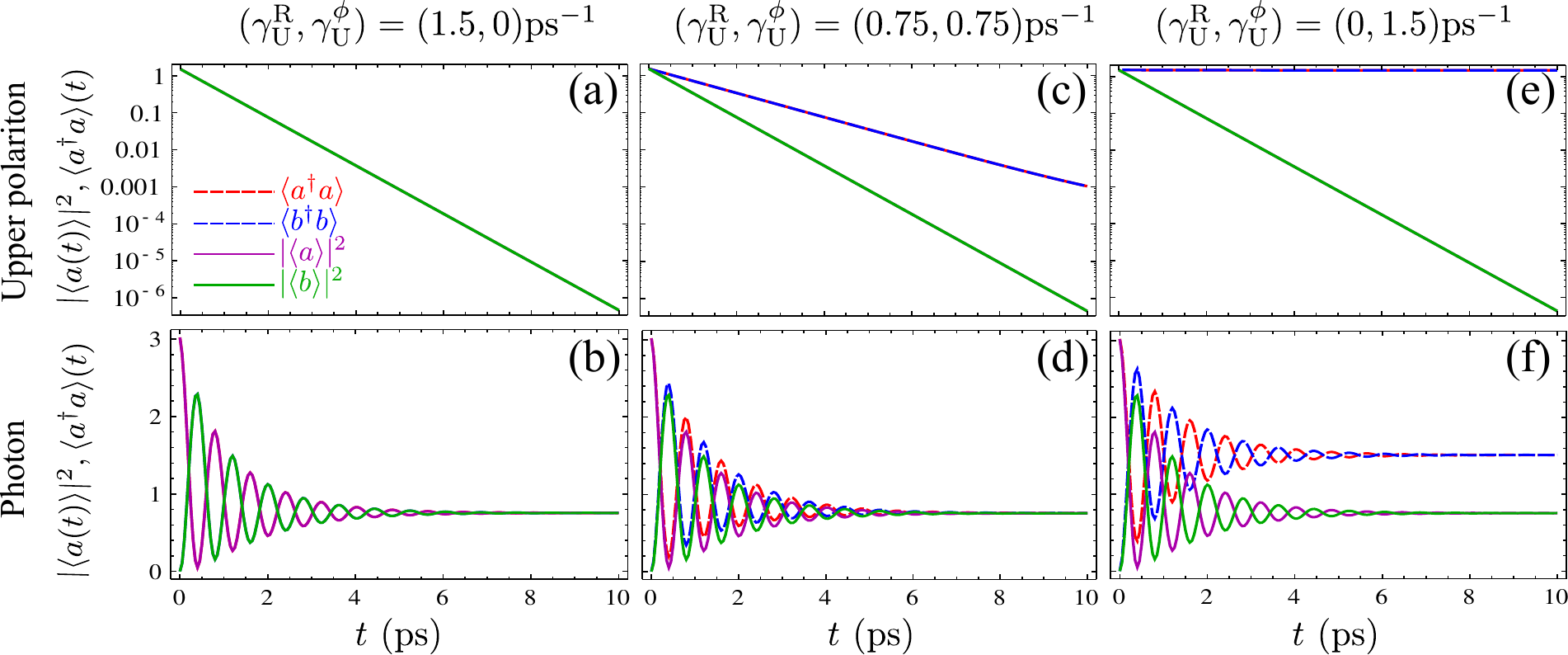}
  \caption{(Color online) Effect of dephasing on the polariton
    population. The three cases shown in three columns for two
    different initial conditions (rows) keep the total
    rate~$\gamma_\mathrm{U}=\SI{3/2}{\per\pico\second}$ constant but
    balances radiative and pure dephasing from
    Eq.~(\ref{eq:lunjun30100400CEST2014}) as indicated in the header
    of each column. The evolution of the coherent (solid green \&
    purple) and total, i.e., coherent+incoherent (dashed red \& blue)
    populations of exciton and photon are shown as a function of time,
    starting from a condensate of upper polaritons (upper row) and of
    photons (lower row). The coherent fraction is the same, and looses
    its particle either radiatively or transferred to the incoherent
    fraction.}
  \label{fig:juejul10100806CEST2014}
\end{figure*}

\section{Visualization of the dynamics}
\label{sec:domjun29135759CEST2014}

We now discuss the problem of the visualization of the polariton
dynamics. At a basic level, the problem seems innocuous enough, as it
deals with oscillations, which are essentially captured by their
amplitude and phase. For two fields, this means two complex numbers,
i.e., four variables. Two complex numbers can be mapped onto the Bloch
sphere if dropping their relative phase, which indeed can be done with
no loss of generality. The state of the oscillator at any given
time~$t$ can thus be positioned on a sphere with polar coordinates
defined as:
\begin{equation}
  \label{eq:lunjun30084533CEST2014}
  \theta=\frac{\langle a\rangle}{\sqrt{|\langle a\rangle|^2+|\langle b\rangle|^2}}\quad\mathrm{and}\quad
  \phi=\frac{\langle b\rangle}{\sqrt{|\langle a\rangle|^2+|\langle b\rangle|^2}}\,,
\end{equation}
possibly keeping its radius normalized to one, as we do in this work
for clarity. This is the representation we have adopted whenever we
display a trajectory on the sphere, with always the convention that:
\begin{enumerate}
\addtolength{\itemsep}{-0.5\baselineskip}
\item The north pole corresponds to the Upper Polariton (noted $\ket{\mathrm{U}}$),
\item The south pole corresponds to the Lower Polariton (noted $\ket{\mathrm{L}}$),
\item The right-side point on the equator corresponds to the Cavity Photon (noted $\ket{\mathrm{C}}$),
\item The left-side point on the equator corresponds to the Exciton (noted $\ket{\mathrm{X}}$).
\end{enumerate}
Here it is important to keep in mind that throughout, and regardless
of the notation, this maps a state of the type
Eq.~(\ref{eq:miéjun25182235CEST2014}), specifically,
$\ket{\alpha}\ket{\beta}$, and not a qubit nor a superposition of the
states. The possibility to map to the same sphere either the classical
state of two coherent fields or the quantum state of a qubit
$\cos(\theta/2)\ket{0}\ket{1}+e^{i\phi}\sin(\theta/2)\ket{1}\ket{0}$,
can allow for some classical simulation of a qubit, which may have
some value, but cannot, obviously, substitute for it in an actual
quantum computer. We leave the analysis of quantum oscillations to
another text~\cite{S_unpub_carreno14a} and focus here on the case of
interest for the experiment.

The Bloch sphere representation is the most concise one, but it says
nothing about the incoherent fraction.  In our experiment, it turns
out that the states remain highly coherent throughout, and there is no
need to consider the dynamics of the incoherent part that grows due to
dephasing and incoherent pumping. 
\begin{figure*}[ht]
  \includegraphics[width=.75\linewidth]{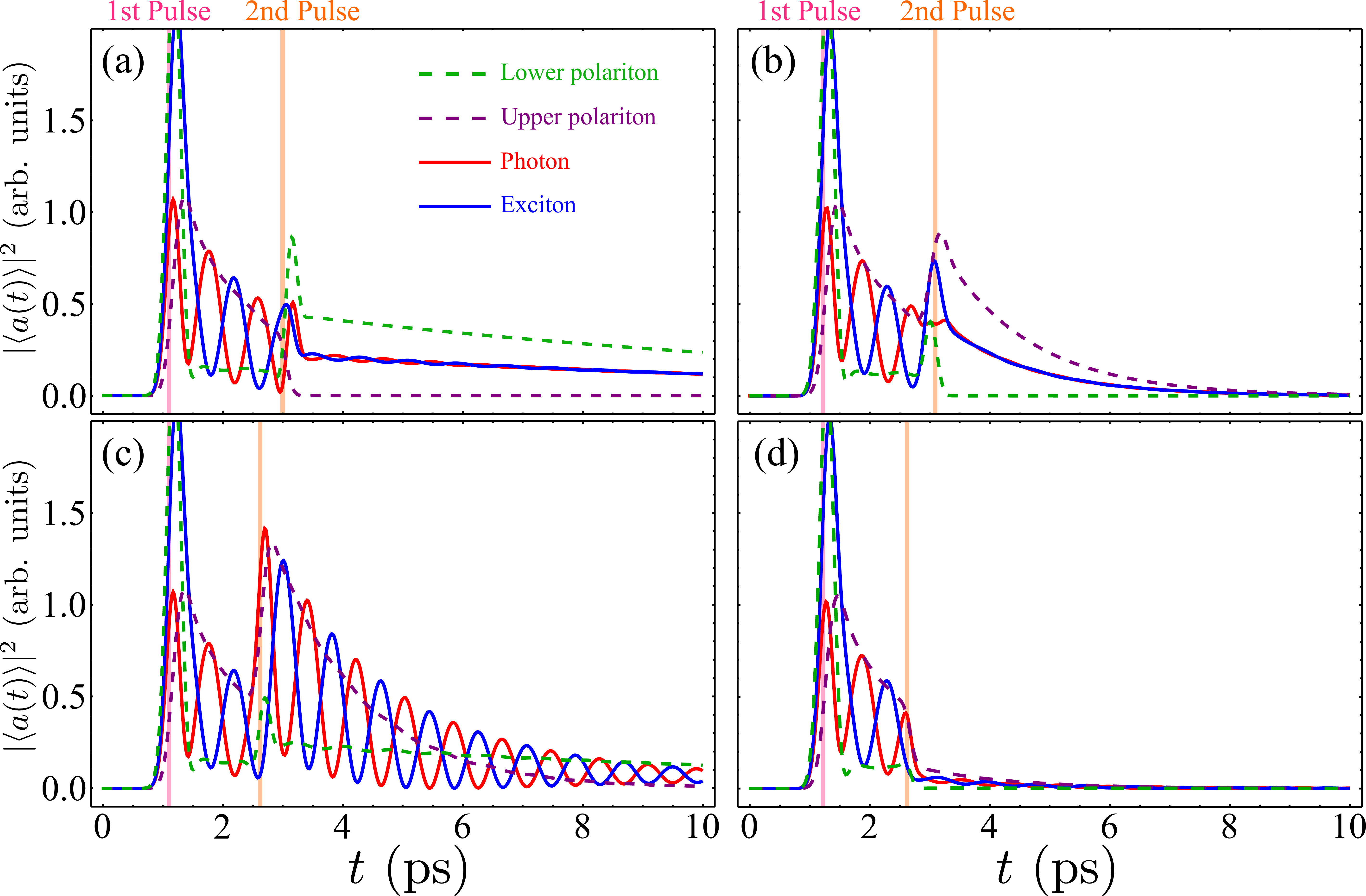}
  \caption{(Color online) Two-pulses experiment (cf.~Fig.~3 of the
    main text) as seen through all the theoretical variables: bare
    states in solid and eigenstates in dashed lines. The
    experimentally available variable is the photon field, in red. The
    three cases correspond to: (a) switching-off of the Rabi
    oscillation by bringing the state into a lower-polariton
    condensate, (b) revival of the oscillations and (c) annihilation
    of the field.}
  \label{fig:lunjun30090805CEST2014}
\end{figure*}
However, to be comprehensive, since we still need the relative phase
information of both fields, and to put the current results in
perspective with future ones with smaller number of particles, where
more general states can be realized, we now discuss a more
comprehensive picture, namely, the Husimi
representation~\cite{S_husimi40a}. It is given for two coherent fields
by:
\begin{equation}
  \label{eq:lunjun30093241CEST2014}
  Q(\alpha,\beta)=\bra{\alpha\beta}\rho\ket{\alpha\beta}\,,
\end{equation}
and can be represented in its reduced form for each fields in two
panels, juxtaposing $Q(\alpha)=\int Q(\alpha,\beta)\,d\beta$ and
$Q(\beta)=\int Q(\alpha,\beta)\,d\alpha$. When it is a product of
coherent states, the factorization is exact and no information is
lost, in which case it is equivalent to the complex phase
representation that we have used in the text, cf.~Fig.~3(a), merely
replacing the point by a Gaussian cloud of mean square deviation~$1/2$
and whose amplitude and phase are otherwise given by the population
and phase of the oscillator. This is arguably the most convenient way
to visualize the polariton dynamics of normal-mode coupling,
particularly if it can be animated.  In the supplementary video
\texttt{II-RabiOscillations.mp4}~\cite{S_movieII}, we provide the same
dynamics as Fig.~3(a), namely, the sequence of pulses that bring the
systems into the successive states:
\begin{enumerate}
\addtolength{\itemsep}{-0.5\baselineskip}
\item vacuum,
\item excitation of a photon condensate,
\item transfer to an upper polariton condensate,
\item transfer to an exciton condensate,
\item transfer to a lower polariton condensate,
\item annihilation and return to the vacuum.
\end{enumerate}

This is the theoretical, ideal version of what the experiment realizes
with one operation at a time, since the time-window available to us is
not currently large enough to chain up various pulses. However the
proof of principle has been fully demonstrated.

Although in our case the fitting implies that the system remains
highly coherent at all times, before we turn to the details of the
dynamics in this case, we contrast it with cases where dephasing plays
a more important role. In Fig.~\ref{fig:juejul10100806CEST2014}, we
show the dynamics of the system prepared in the upper polariton state
(which is the state that suffers the most from dephasing) for three
different balancing of the total dephasing rate~$\gamma_\mathrm{U}$
given by Eq.~(\ref{eq:lunjun30100400CEST2014})---that is the parameter
accessible to the experiment---into radiative upper polariton
decay~$\gamma_\mathrm{U}^\mathrm{R}$ and pure dephasing of the upper
polariton~$\gamma_\mathrm{U}^\phi$.  The calculations, done in this
case with the master equation, involved only a few particles, so that
the effect of dephasing be noticeable. In the experiment, where it is
estimated as orders of magnitudes higher, the impact would not be
significant on the timescales involved. To keep the discussion as
simple as possible, we consider upper polariton dephasing only and as
initial condition, coherent states of upper polaritons or of photons,
each with 3 particles at~$t=0$. One can see how the breakdown of
polariton dephasing into radiative decay and pure dephasing results in
different evolution of the quantum state, although the coherent
fraction (solid line) remains the same, being dependent only on the
total dephasing rate~(\ref{eq:lunjun30100400CEST2014}). Only in the
case where there is some level of pure dephasing do the two cases
differ. Compare in particular the first and third columns, where the
polaritons are either lost (first column) or transferred to the
incoherent fraction (third column). In the latter case, the population
remain constant when starting as a polariton or becomes so when the
polariton fraction has vanished to leave a fully incoherent mixture.
The movie corresponding to the case~(c) in the Husimi representation
is provided in the Supplementary Material as
\texttt{III-RabiDephasing.avi}~\cite{S_movieIII}. The Gaussian cloud
spreads into a ring as it rotates, corresponding to the washing out of
the phase. This together with the various modes of
oscillations~\cite{S_movieI} give a hint as to what the complete
polariton dynamics look like, when adding to such dephasing also the
radiative decay and Rabi oscillations.


\section{Two-pulses control}

As we have seen in the previous Section, one value of the theory is to
unveil the full dynamics and gain access to variables not reachable by
the experiment. In Fig.~\ref{fig:lunjun30090805CEST2014}, we show the
representation for the two-pulses dynamics of the experiment
(cf.~Fig.~3 of the main text where the experimental points are also
shown), here supplemented with the observables made accessible by the
theory, namely also $|\langle b(t)\rangle|^2$ (solid blue) the
amplitude squared of the exciton condensate and $|\langle
u(t)\rangle|^2$ the upper polariton (dashed purple) and $|\langle
l(t)\rangle|^2$ the lower polariton (dashed green).

This representation gives another look at the physics already
discussed. In the first case, panel~(a), the state is brought from an
oscillating exciton-photon dynamics into a lower polariton, therefore
switching-off the oscillations. The lifetime is also changed as a
result.  In the second case, panel~(b), the transfer is to the upper
polariton instead (a case not achieved in the experiment). The
phenomenology is the same but with a faster decay rate. In case~(c),
the oscillation is revived and carries on Rabi-oscillating beyond our
experimental window. This shows in particular how the upper polariton
is, again, the limiting factor, although, from the fit, it appears it
is easily set in motion and populated, more than the lower polariton
itself. Since it decays so quickly, ultimately the system is driven
into a lower polariton state only, regardless of its excitation at
initial time. Here again, we must stress that the decay of the
contrast of oscillations with time is therefore not due to dephasing,
but because upper polaritons are lost at a greater rate, making the
system increasingly lower-polaritonic, which is a non-oscillating
state.  In the last panel, (d), we provide the clear observation that
the field is annihilated altogether in this configuration where both
the Rabi and the optical phases of the second pulses are in
opposition, which is not apparent on the Bloch sphere that enforces
the normalization.



\section{Fitting of the data}
\label{sec:domjun29114211CEST2014}

\begin{figure}
  \includegraphics[width=1\linewidth]{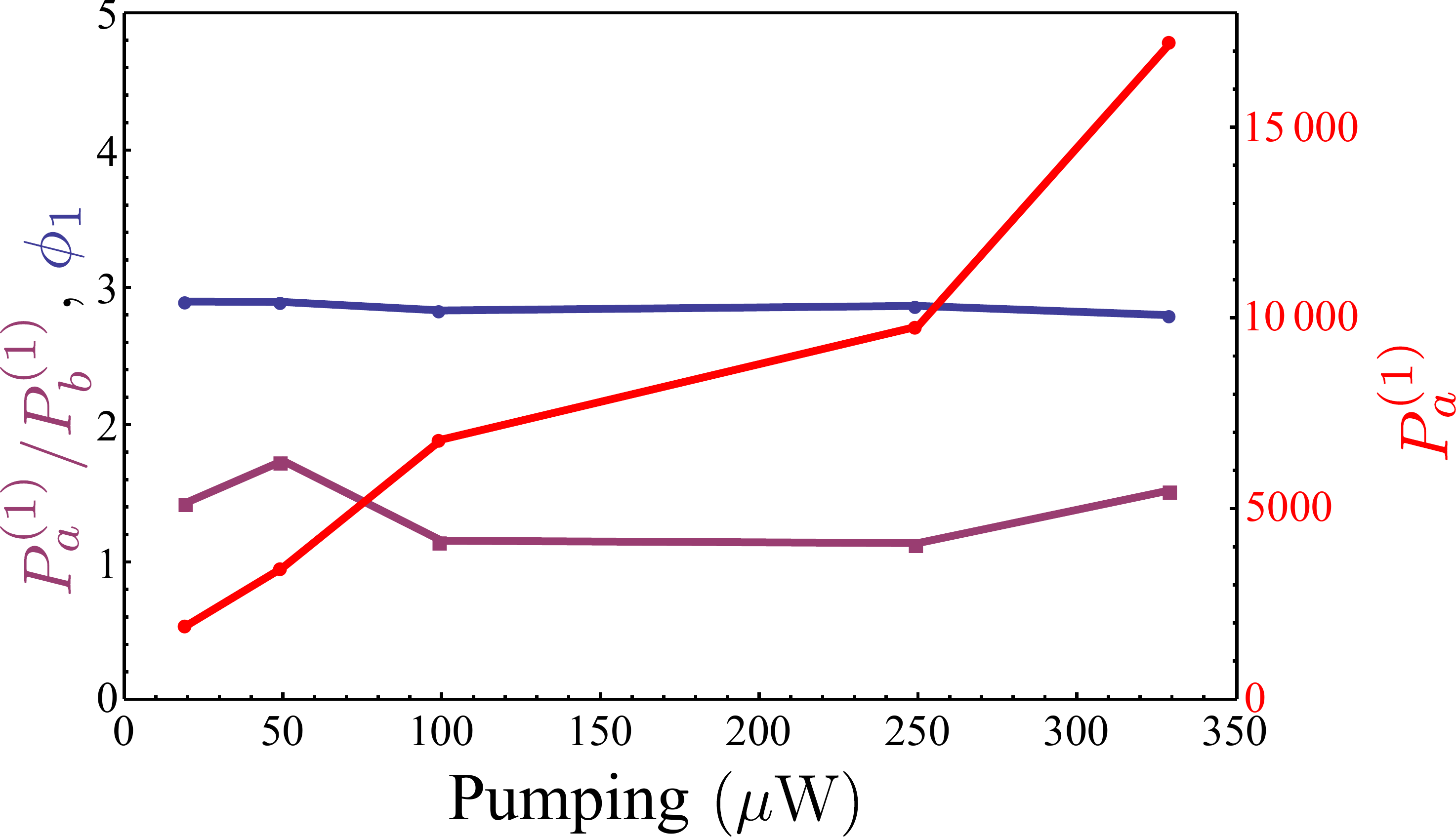}
  \caption{(Color online) Best fitting numerical value for the pumping
    parameters $P_a^{(1)}$ (red), right axis, and the ratio
    $P_a^{(1)}/P_b^{(1)}$, roughly constant, and relative phase~$\phi_A$
    (left axis), in the pumping series.}
  \label{fig:SM2}
\end{figure}

\begin{figure}[h]
  \includegraphics[width=1\linewidth]{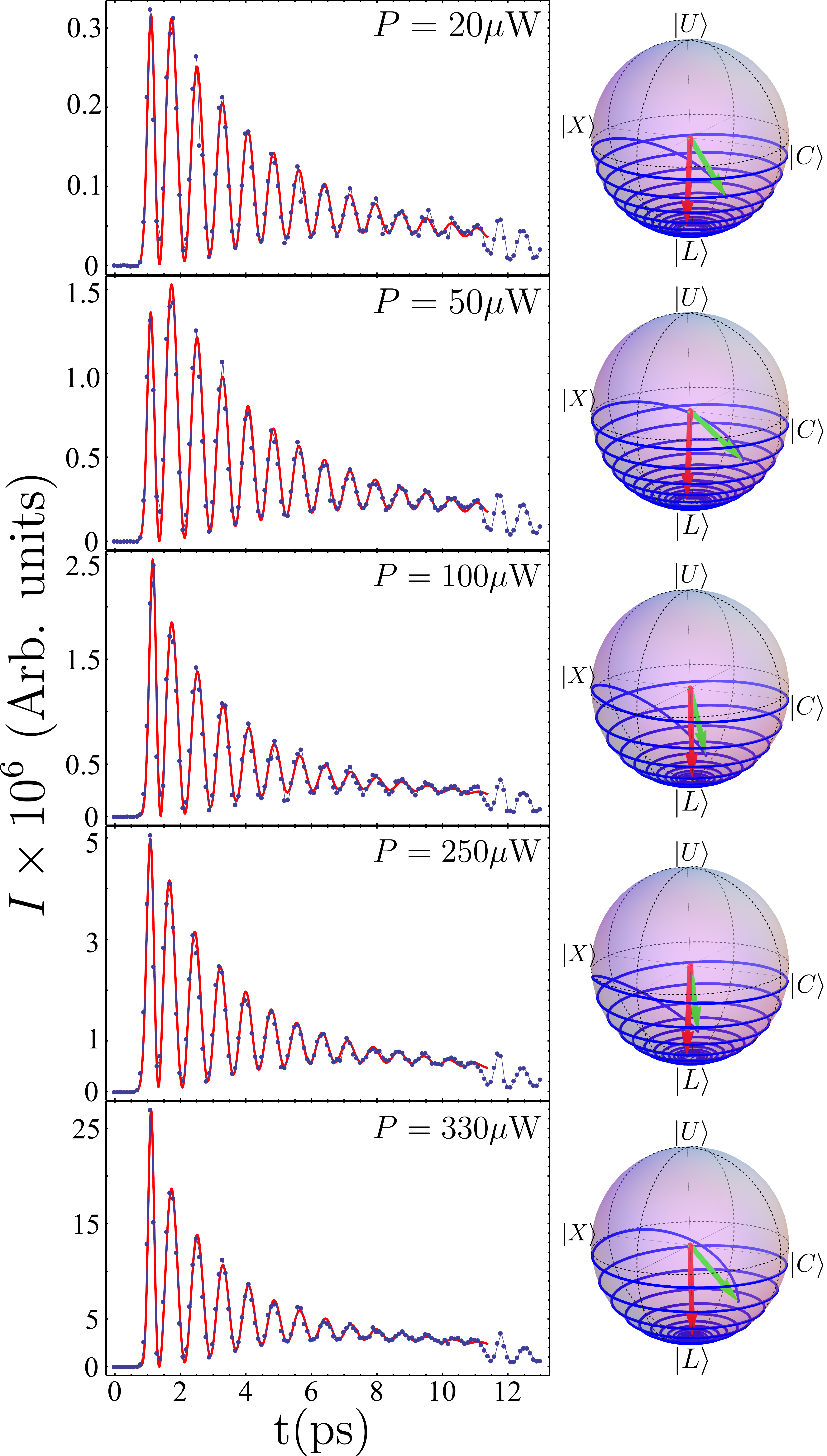}
  \caption{(Color online) Power serie experiment. The evolution of the
    system is presented for different powers. On the graphs, the blue
    points are the experimental datas, linked by straight lines to
    guid the eye. The red curves are obtained with the fitting
    process. The corresponding quantum state is represented on the
    Bloch Sphere, the initial state being represented by the green
    arrow and the final state by the red one.}
  \label{fig:SM1}
\end{figure}

In this last Section, we provide additional material on the fitting of
the experimental data by the theory, i.e., by
Eqs.~(\ref{eq:viejun27093513CEST2014}).

Parameters have been optimized through a multi-pass fitting procedure
that first adjust all parameters and then constrain the system
parameters, i.e., those specific to the sample---which are listed in
Table~\ref{tab:1} for the $\omega_\mathrm{L}$ series provided in the
text---and fit only over those of the pulse in a global-fitting
procedure over various experiments. The fit is not sensitive to the
exciton lifetime as long as it is very large and we have fixed it to
\SI{1}{\nano\second}, typical of a reservoir exciton. This provides an
essentially perfect fit to the data with very few completely free
parameters, most of them being kept fixed in the global fit, and those
of the pulse varying as dictated by the experiment
($\omega_\mathrm{L}$). This leaves only $P_c^{(i)}$ and $\phi_c$,
$c=a$, $b$ and~$i=1$, $2$, as the truly free parameters, providing
useful information on how the laser couples to the microcavity.

\begin{table}
  \begin{tabular}{ l l r }
    \hhline{===}
    Parameters & Physical Meaning & Best Fit \\
    \hline
    $g$ & Coupling Strength & 2.65 $\textrm{ps}^{-1}$ \\
    $\gamma_a$ & Cavity decay rate & 0.2 $\textrm{ps}^{-1}$ \\
    $\gamma_b$ & Exciton decay rate & 0.001 $\textrm{ps}^{-1}$ \\
    $\gamma_\mathrm{U}$ & Upper polariton dephasing rate & 0.43 $\textrm{ps}^{-1}$ \\
    $P_b$ & Exciton reservoir pumping rate & 0.11 $\textrm{ps}^{-1}$ \\
    $\gamma_{P_b}$ & Exciton reservoir decay rate & 0.01 $\textrm{ps}^{-1}$ \\ 
    \hhline{===}
  \end{tabular}
  \caption{System parameters and their best fit values for the energy series experiment (varying $\omega_\mathrm{L}$, cf.~Fig.~2 of the main text). The same parameters apply for the pumping series, only with $\gamma_\mathrm{U}=1/\SI{0.41}{\pico\second}$}
  \label{tab:1}
\end{table}

In addition to the laser energy series, presented in the main text, we
have also studied the pumping dependence.  The results we have
reported lie in the low excitation regime to retain the linear
features, and we are going to show the stability of such a regime over
one order of magnitude pumping. At higher powers, deviations appear
that will be analyzed separately. At very high powers, a new physics
altogether emerge with spectacular phenomenology such as the
observation of a long-living, ultrasharp
backjet~\cite{S_arXiv_dominici13a}.  Figure~\ref{fig:SM1} shows the data
and its fit by the model as the power of the laser is tuned from
$\SI{20}{\micro\watt}$ to $\SI{330}{\micro\watt}$ while the excitation
energy is kept constant at $\omega_\mathrm{L}=\SI{829}{\nano\meter}$.
The same multi-pass fitting procedure was used for the power serie and
provides again an essentially perfect fit of the experiment. The
corresponding parameters, summarized in the Table~\ref{tab:1}, present
an excellent agreement with the ones obtained previously for the
energy serie (cf.~see Table~1 in the main text).  Figure~\ref{fig:SM2}
displays the numerical values for $P_a^{(1)}$ (red) and $\phi_A$
(blue) as well as the ratio~$P_a^{(1)}/P_b^{(1)}$ (purple).  The
fitting values increase roughly linearly with pumping, as expected
since this is precisely the variable that is tuned
experimentally. This also indicates how the laser couples to the
system. This is an information that is not easily obtained and from
which we gather that the laser couples in equal part to the photon
than to the exciton, both being almost in optical antiphase.  That the
coupling of the laser to the microcavity is not only through the
cavity but also to the exciton is important for coherent control in
particular but may have implications in other aspect of polariton
physics. The coupling is found to be independent of pumping, as shown
by the constant ratio and phase, which leads to an initial state on
the sphere in the same vicinity, see the green arrows on the different
sphere in Fig.~\ref{fig:SM2}.  This also shows that the reservoir
grows linearly with pumping, since the rate is fixed.  The shape of
the Rabi oscillations after the pulse remains independent of the
pumping, only scaling in total particle numbers. The apparent
different contrast with pumping is an artifact due to the larger
contribution of the pulse at higher pumping, dwarfing the rest of the
free dynamics.


\begin{thebibliography}{45}
\expandafter\ifx\csname natexlab\endcsname\relax\def\natexlab#1{#1}\fi
\expandafter\ifx\csname bibnamefont\endcsname\relax
  \def\bibnamefont#1{#1}\fi
\expandafter\ifx\csname bibfnamefont\endcsname\relax
  \def\bibfnamefont#1{#1}\fi
\expandafter\ifx\csname citenamefont\endcsname\relax
  \def\citenamefont#1{#1}\fi
\expandafter\ifx\csname url\endcsname\relax
  \def\url#1{\texttt{#1}}\fi
\expandafter\ifx\csname urlprefix\endcsname\relax\def\urlprefix{URL }\fi
\providecommand{\bibinfo}[2]{#2}
\providecommand{\eprint}[2][]{\url{#2}}

\bibitem[{\citenamefont{Rabi}(1937)}]{rabi37a}
\bibinfo{author}{\bibfnamefont{I.~I.} \bibnamefont{Rabi}},
  \bibinfo{journal}{Phys. Rev.} \textbf{\bibinfo{volume}{51}},
  \bibinfo{pages}{652} (\bibinfo{year}{1937}).

\bibitem[{\citenamefont{Schumacher}(1995)}]{schumacher95a}
\bibinfo{author}{\bibfnamefont{B.}~\bibnamefont{Schumacher}},
  \bibinfo{journal}{Phys. Rev. A} \textbf{\bibinfo{volume}{51}},
  \bibinfo{pages}{2738} (\bibinfo{year}{1995}).

\bibitem[{\citenamefont{Nielsen and Chuang}(2000)}]{nielsen_book00a}
\bibinfo{author}{\bibfnamefont{M.~A.} \bibnamefont{Nielsen}} \bibnamefont{and}
  \bibinfo{author}{\bibfnamefont{I.~L.} \bibnamefont{Chuang}},
  \emph{\bibinfo{title}{Quantum computation and quantum information}}
  (\bibinfo{publisher}{Cambridge University Press}, \bibinfo{year}{2000}).

\bibitem[{\citenamefont{Spreeuw and Woerdman}(1993)}]{spreeuw93a}
\bibinfo{author}{\bibfnamefont{R.}~\bibnamefont{Spreeuw}} \bibnamefont{and}
  \bibinfo{author}{\bibfnamefont{J.}~\bibnamefont{Woerdman}},
  \bibinfo{journal}{Progress in Optics} \textbf{\bibinfo{volume}{31}},
  \bibinfo{pages}{263} (\bibinfo{year}{1993}).

\bibitem[{\citenamefont{Faust et~al.}(2013)\citenamefont{Faust, Rieger,
  Seitner, Kotthaus, and Weig}}]{faust13a}
\bibinfo{author}{\bibfnamefont{T.}~\bibnamefont{Faust}},
  \bibinfo{author}{\bibfnamefont{J.}~\bibnamefont{Rieger}},
  \bibinfo{author}{\bibfnamefont{M.~J.} \bibnamefont{Seitner}},
  \bibinfo{author}{\bibfnamefont{J.~P.} \bibnamefont{Kotthaus}},
  \bibnamefont{and} \bibinfo{author}{\bibfnamefont{E.~M.} \bibnamefont{Weig}},
  \bibinfo{journal}{Nat. Phys.} \textbf{\bibinfo{volume}{9}},
  \bibinfo{pages}{485} (\bibinfo{year}{2013}).

\bibitem[{\citenamefont{Zhu et~al.}(1990)\citenamefont{Zhu, Gauthier, Morin,
  Wu, Carmichael, and Mossberg}}]{zhu90a}
\bibinfo{author}{\bibfnamefont{Y.}~\bibnamefont{Zhu}},
  \bibinfo{author}{\bibfnamefont{D.~J.} \bibnamefont{Gauthier}},
  \bibinfo{author}{\bibfnamefont{S.~E.} \bibnamefont{Morin}},
  \bibinfo{author}{\bibfnamefont{Q.}~\bibnamefont{Wu}},
  \bibinfo{author}{\bibfnamefont{H.~J.} \bibnamefont{Carmichael}},
  \bibnamefont{and} \bibinfo{author}{\bibfnamefont{T.~W.}
  \bibnamefont{Mossberg}}, \bibinfo{journal}{Phys. Rev. Lett.}
  \textbf{\bibinfo{volume}{64}}, \bibinfo{pages}{2499} (\bibinfo{year}{1990}).

\bibitem[{\citenamefont{Khitrova et~al.}(2006)\citenamefont{Khitrova, Gibbs,
  Kira, Koch, and Scherer}}]{khitrova06a}
\bibinfo{author}{\bibfnamefont{G.}~\bibnamefont{Khitrova}},
  \bibinfo{author}{\bibfnamefont{H.~M.} \bibnamefont{Gibbs}},
  \bibinfo{author}{\bibfnamefont{M.}~\bibnamefont{Kira}},
  \bibinfo{author}{\bibfnamefont{S.~W.} \bibnamefont{Koch}}, \bibnamefont{and}
  \bibinfo{author}{\bibfnamefont{A.}~\bibnamefont{Scherer}},
  \bibinfo{journal}{Nat. Phys.} \textbf{\bibinfo{volume}{2}},
  \bibinfo{pages}{81} (\bibinfo{year}{2006}).

\bibitem[{\citenamefont{Matthews et~al.}(1999)\citenamefont{Matthews, Anderson,
  Haljan, Hall, Holland, Williams, Wieman, and Cornell}}]{matthews99b}
\bibinfo{author}{\bibfnamefont{M.~R.} \bibnamefont{Matthews}},
  \bibinfo{author}{\bibfnamefont{B.~P.} \bibnamefont{Anderson}},
  \bibinfo{author}{\bibfnamefont{P.~C.} \bibnamefont{Haljan}},
  \bibinfo{author}{\bibfnamefont{D.~S.} \bibnamefont{Hall}},
  \bibinfo{author}{\bibfnamefont{M.~J.} \bibnamefont{Holland}},
  \bibinfo{author}{\bibfnamefont{J.~E.} \bibnamefont{Williams}},
  \bibinfo{author}{\bibfnamefont{C.~E.} \bibnamefont{Wieman}},
  \bibnamefont{and} \bibinfo{author}{\bibfnamefont{E.~A.}
  \bibnamefont{Cornell}}, \bibinfo{journal}{Phys. Rev. Lett.}
  \textbf{\bibinfo{volume}{83}}, \bibinfo{pages}{3358} (\bibinfo{year}{1999}).

\bibitem[{\citenamefont{Vasa et~al.}(2013)\citenamefont{Vasa, Wang, Pomraenke,
  Lammers, Maiuri, Manzoni, Cerullo, and Lienau}}]{vasa13a}
\bibinfo{author}{\bibfnamefont{P.}~\bibnamefont{Vasa}},
  \bibinfo{author}{\bibfnamefont{W.}~\bibnamefont{Wang}},
  \bibinfo{author}{\bibfnamefont{R.}~\bibnamefont{Pomraenke}},
  \bibinfo{author}{\bibfnamefont{M.}~\bibnamefont{Lammers}},
  \bibinfo{author}{\bibfnamefont{M.}~\bibnamefont{Maiuri}},
  \bibinfo{author}{\bibfnamefont{C.}~\bibnamefont{Manzoni}},
  \bibinfo{author}{\bibfnamefont{G.}~\bibnamefont{Cerullo}}, \bibnamefont{and}
  \bibinfo{author}{\bibfnamefont{C.}~\bibnamefont{Lienau}},
  \bibinfo{journal}{Nat. Photon.} \textbf{\bibinfo{volume}{7}},
  \bibinfo{pages}{128} (\bibinfo{year}{2013}).

\bibitem[{\citenamefont{Spreeuw}(1998)}]{spreeuw98a}
\bibinfo{author}{\bibfnamefont{R.~J.~C.} \bibnamefont{Spreeuw}},
  \bibinfo{journal}{Found. Phys.} \textbf{\bibinfo{volume}{28}},
  \bibinfo{pages}{361} (\bibinfo{year}{1998}).

\bibitem[{\citenamefont{Spreeuw}(2001)}]{spreeuw01a}
\bibinfo{author}{\bibfnamefont{R.~J.~C.} \bibnamefont{Spreeuw}},
  \bibinfo{journal}{Phys. Rev. A} \textbf{\bibinfo{volume}{63}},
  \bibinfo{pages}{062302} (\bibinfo{year}{2001}).

\bibitem[{\citenamefont{Dragoman and Dragoman}(2004)}]{dragoman_book04a}
\bibinfo{author}{\bibfnamefont{D.}~\bibnamefont{Dragoman}} \bibnamefont{and}
  \bibinfo{author}{\bibfnamefont{M.}~\bibnamefont{Dragoman}},
  \emph{\bibinfo{title}{Quantum-Classical Analogies}}, The Frontiers Collection
  (\bibinfo{publisher}{Springer}, \bibinfo{year}{2004}).

\bibitem[{\citenamefont{Spreeuw et~al.}(1990)\citenamefont{Spreeuw, van Druten,
  Beijersbergen, Eliel, and Woerdman}}]{spreeuw90a}
\bibinfo{author}{\bibfnamefont{R.~J.~C.} \bibnamefont{Spreeuw}},
  \bibinfo{author}{\bibfnamefont{N.~J.} \bibnamefont{van Druten}},
  \bibinfo{author}{\bibfnamefont{M.~W.} \bibnamefont{Beijersbergen}},
  \bibinfo{author}{\bibfnamefont{E.~R.} \bibnamefont{Eliel}}, \bibnamefont{and}
  \bibinfo{author}{\bibfnamefont{J.~P.} \bibnamefont{Woerdman}},
  \bibinfo{journal}{Phys. Rev. Lett.} \textbf{\bibinfo{volume}{65}},
  \bibinfo{pages}{2642} (\bibinfo{year}{1990}).

\bibitem[{\citenamefont{Okamoto et~al.}(2013)\citenamefont{Okamoto, Gourgout,
  Chang, Onomitsu, Mahboob, Chang, and Yamaguchi}}]{okamoto13a}
\bibinfo{author}{\bibfnamefont{H.}~\bibnamefont{Okamoto}},
  \bibinfo{author}{\bibfnamefont{A.}~\bibnamefont{Gourgout}},
  \bibinfo{author}{\bibfnamefont{C.-Y.} \bibnamefont{Chang}},
  \bibinfo{author}{\bibfnamefont{K.}~\bibnamefont{Onomitsu}},
  \bibinfo{author}{\bibfnamefont{I.}~\bibnamefont{Mahboob}},
  \bibinfo{author}{\bibfnamefont{E.~Y.} \bibnamefont{Chang}}, \bibnamefont{and}
  \bibinfo{author}{\bibfnamefont{H.}~\bibnamefont{Yamaguchi}},
  \bibinfo{journal}{Nat. Phys.} \textbf{\bibinfo{volume}{9}},
  \bibinfo{pages}{480} (\bibinfo{year}{2013}).

\bibitem[{\citenamefont{Hennessy et~al.}(2007)\citenamefont{Hennessy, Badolato,
  Winger, Gerace, Atature, Gulde, {F\u alt}, Hu, and {\u
  Imamo\=glu}}}]{hennessy07a}
\bibinfo{author}{\bibfnamefont{K.}~\bibnamefont{Hennessy}},
  \bibinfo{author}{\bibfnamefont{A.}~\bibnamefont{Badolato}},
  \bibinfo{author}{\bibfnamefont{M.}~\bibnamefont{Winger}},
  \bibinfo{author}{\bibfnamefont{D.}~\bibnamefont{Gerace}},
  \bibinfo{author}{\bibfnamefont{M.}~\bibnamefont{Atature}},
  \bibinfo{author}{\bibfnamefont{S.}~\bibnamefont{Gulde}},
  \bibinfo{author}{\bibfnamefont{S.}~\bibnamefont{{F\u alt}}},
  \bibinfo{author}{\bibfnamefont{E.~L.} \bibnamefont{Hu}}, \bibnamefont{and}
  \bibinfo{author}{\bibfnamefont{A.}~\bibnamefont{{\u Imamo\=glu}}},
  \bibinfo{journal}{Nature} \textbf{\bibinfo{volume}{445}},
  \bibinfo{pages}{896} (\bibinfo{year}{2007}).

\bibitem[{\citenamefont{Weisbuch et~al.}(1992)\citenamefont{Weisbuch, Nishioka,
  Ishikawa, and Arakawa}}]{weisbuch92a}
\bibinfo{author}{\bibfnamefont{C.}~\bibnamefont{Weisbuch}},
  \bibinfo{author}{\bibfnamefont{M.}~\bibnamefont{Nishioka}},
  \bibinfo{author}{\bibfnamefont{A.}~\bibnamefont{Ishikawa}}, \bibnamefont{and}
  \bibinfo{author}{\bibfnamefont{Y.}~\bibnamefont{Arakawa}},
  \bibinfo{journal}{Phys. Rev. Lett.} \textbf{\bibinfo{volume}{69}},
  \bibinfo{pages}{3314} (\bibinfo{year}{1992}).

\bibitem[{\citenamefont{Kavokin et~al.}(2011)\citenamefont{Kavokin, Baumberg,
  Malpuech, and Laussy}}]{kavokin_book11a}
\bibinfo{author}{\bibfnamefont{A.}~\bibnamefont{Kavokin}},
  \bibinfo{author}{\bibfnamefont{J.~J.} \bibnamefont{Baumberg}},
  \bibinfo{author}{\bibfnamefont{G.}~\bibnamefont{Malpuech}}, \bibnamefont{and}
  \bibinfo{author}{\bibfnamefont{F.~P.} \bibnamefont{Laussy}},
  \emph{\bibinfo{title}{Microcavities}} (\bibinfo{publisher}{Oxford University
  Press}, \bibinfo{year}{2011}), \bibinfo{edition}{2nd} ed.

\bibitem[{\citenamefont{Carusotto and Ciuti}(2013)}]{carusotto13a}
\bibinfo{author}{\bibfnamefont{I.}~\bibnamefont{Carusotto}} \bibnamefont{and}
  \bibinfo{author}{\bibfnamefont{C.}~\bibnamefont{Ciuti}},
  \bibinfo{journal}{Rev. Mod. Phys.} \textbf{\bibinfo{volume}{85}},
  \bibinfo{pages}{299} (\bibinfo{year}{2013}).

\bibitem[{\citenamefont{Kasprzak et~al.}(2006)\citenamefont{Kasprzak, Richard,
  Kundermann, Baas, Jeambrun, Keeling, Marchetti, Szymanska, Andr\'e, Staehli
  et~al.}}]{kasprzak06a}
\bibinfo{author}{\bibfnamefont{J.}~\bibnamefont{Kasprzak}},
  \bibinfo{author}{\bibfnamefont{M.}~\bibnamefont{Richard}},
  \bibinfo{author}{\bibfnamefont{S.}~\bibnamefont{Kundermann}},
  \bibinfo{author}{\bibfnamefont{A.}~\bibnamefont{Baas}},
  \bibinfo{author}{\bibfnamefont{P.}~\bibnamefont{Jeambrun}},
  \bibinfo{author}{\bibfnamefont{J.~M.~J.} \bibnamefont{Keeling}},
  \bibinfo{author}{\bibfnamefont{F.~M.} \bibnamefont{Marchetti}},
  \bibinfo{author}{\bibfnamefont{M.~H.} \bibnamefont{Szymanska}},
  \bibinfo{author}{\bibfnamefont{R.}~\bibnamefont{Andr\'e}},
  \bibinfo{author}{\bibfnamefont{J.~L.} \bibnamefont{Staehli}},
  \bibnamefont{et~al.}, \bibinfo{journal}{Nature}
  \textbf{\bibinfo{volume}{443}}, \bibinfo{pages}{409} (\bibinfo{year}{2006}).

\bibitem[{\citenamefont{Amo et~al.}(2009{\natexlab{a}})\citenamefont{Amo,
  Sanvitto, Laussy, Ballarini, del Valle, Martin, Lema\^itre, Bloch,
  Krizhanovskii, Skolnick et~al.}}]{amo09a}
\bibinfo{author}{\bibfnamefont{A.}~\bibnamefont{Amo}},
  \bibinfo{author}{\bibfnamefont{D.}~\bibnamefont{Sanvitto}},
  \bibinfo{author}{\bibfnamefont{F.~P.} \bibnamefont{Laussy}},
  \bibinfo{author}{\bibfnamefont{D.}~\bibnamefont{Ballarini}},
  \bibinfo{author}{\bibfnamefont{E.}~\bibnamefont{del Valle}},
  \bibinfo{author}{\bibfnamefont{M.~D.} \bibnamefont{Martin}},
  \bibinfo{author}{\bibfnamefont{A.}~\bibnamefont{Lema\^itre}},
  \bibinfo{author}{\bibfnamefont{J.}~\bibnamefont{Bloch}},
  \bibinfo{author}{\bibfnamefont{D.~N.} \bibnamefont{Krizhanovskii}},
  \bibinfo{author}{\bibfnamefont{M.~S.} \bibnamefont{Skolnick}},
  \bibnamefont{et~al.}, \bibinfo{journal}{Nature}
  \textbf{\bibinfo{volume}{457}}, \bibinfo{pages}{291}
  (\bibinfo{year}{2009}{\natexlab{a}}).

\bibitem[{\citenamefont{Amo et~al.}(2009{\natexlab{b}})\citenamefont{Amo,
  Lefr\`ere, Pigeon, Adrados, Ciuti, Carusotto, Houdr\'e, Giacobino, and
  Bramati}}]{amo09b}
\bibinfo{author}{\bibfnamefont{A.}~\bibnamefont{Amo}},
  \bibinfo{author}{\bibfnamefont{J.}~\bibnamefont{Lefr\`ere}},
  \bibinfo{author}{\bibfnamefont{S.}~\bibnamefont{Pigeon}},
  \bibinfo{author}{\bibfnamefont{C.}~\bibnamefont{Adrados}},
  \bibinfo{author}{\bibfnamefont{C.}~\bibnamefont{Ciuti}},
  \bibinfo{author}{\bibfnamefont{I.}~\bibnamefont{Carusotto}},
  \bibinfo{author}{\bibfnamefont{R.}~\bibnamefont{Houdr\'e}},
  \bibinfo{author}{\bibfnamefont{E.}~\bibnamefont{Giacobino}},
  \bibnamefont{and} \bibinfo{author}{\bibfnamefont{A.}~\bibnamefont{Bramati}},
  \bibinfo{journal}{Nat. Phys.} \textbf{\bibinfo{volume}{5}},
  \bibinfo{pages}{805} (\bibinfo{year}{2009}{\natexlab{b}}).

\bibitem[{\citenamefont{Liew et~al.}(2008)\citenamefont{Liew, Kavokin, and
  Shelykh}}]{liew08a}
\bibinfo{author}{\bibfnamefont{T.~C.~H.} \bibnamefont{Liew}},
  \bibinfo{author}{\bibfnamefont{A.~V.} \bibnamefont{Kavokin}},
  \bibnamefont{and} \bibinfo{author}{\bibfnamefont{I.~A.}
  \bibnamefont{Shelykh}}, \bibinfo{journal}{Phys. Rev. Lett.}
  \textbf{\bibinfo{volume}{101}}, \bibinfo{pages}{016402}
  (\bibinfo{year}{2008}).

\bibitem[{\citenamefont{Amo et~al.}(2010)\citenamefont{Amo, Liew, Adrados,
  Houdr\'e, Giacobino, Kavokin, and Bramati}}]{amo10a}
\bibinfo{author}{\bibfnamefont{A.}~\bibnamefont{Amo}},
  \bibinfo{author}{\bibfnamefont{T.~C.~H.} \bibnamefont{Liew}},
  \bibinfo{author}{\bibfnamefont{C.}~\bibnamefont{Adrados}},
  \bibinfo{author}{\bibfnamefont{R.}~\bibnamefont{Houdr\'e}},
  \bibinfo{author}{\bibfnamefont{E.}~\bibnamefont{Giacobino}},
  \bibinfo{author}{\bibfnamefont{A.~V.} \bibnamefont{Kavokin}},
  \bibnamefont{and} \bibinfo{author}{\bibfnamefont{A.}~\bibnamefont{Bramati}},
  \bibinfo{journal}{Nat. Photon.} \textbf{\bibinfo{volume}{4}},
  \bibinfo{pages}{361} (\bibinfo{year}{2010}).

\bibitem[{\citenamefont{Amo et~al.}(2011)\citenamefont{Amo, Pigeon, Sanvitto,
  Sala, Hivet, Carusotto, Pisanello, Lem\'enager, Houdr\'e, Giacobino
  et~al.}}]{amo11a}
\bibinfo{author}{\bibfnamefont{A.}~\bibnamefont{Amo}},
  \bibinfo{author}{\bibfnamefont{S.}~\bibnamefont{Pigeon}},
  \bibinfo{author}{\bibfnamefont{D.}~\bibnamefont{Sanvitto}},
  \bibinfo{author}{\bibfnamefont{V.~G.} \bibnamefont{Sala}},
  \bibinfo{author}{\bibfnamefont{R.}~\bibnamefont{Hivet}},
  \bibinfo{author}{\bibfnamefont{I.}~\bibnamefont{Carusotto}},
  \bibinfo{author}{\bibfnamefont{F.}~\bibnamefont{Pisanello}},
  \bibinfo{author}{\bibfnamefont{G.}~\bibnamefont{Lem\'enager}},
  \bibinfo{author}{\bibfnamefont{R.}~\bibnamefont{Houdr\'e}},
  \bibinfo{author}{\bibfnamefont{E.}~\bibnamefont{Giacobino}},
  \bibnamefont{et~al.}, \bibinfo{journal}{Science}
  \textbf{\bibinfo{volume}{332}}, \bibinfo{pages}{1167} (\bibinfo{year}{2011}).

\bibitem[{\citenamefont{Baumberg et~al.}(2008)\citenamefont{Baumberg, Kavokin,
  Christopoulos, Grundy, Butt\'e, Christmann, Solnyshkov, Malpuech, von
  H\"ogersthal, Feltin et~al.}}]{baumberg08a}
\bibinfo{author}{\bibfnamefont{J.~J.} \bibnamefont{Baumberg}},
  \bibinfo{author}{\bibfnamefont{A.~V.} \bibnamefont{Kavokin}},
  \bibinfo{author}{\bibfnamefont{S.}~\bibnamefont{Christopoulos}},
  \bibinfo{author}{\bibfnamefont{A.~J.~D.} \bibnamefont{Grundy}},
  \bibinfo{author}{\bibfnamefont{R.}~\bibnamefont{Butt\'e}},
  \bibinfo{author}{\bibfnamefont{G.}~\bibnamefont{Christmann}},
  \bibinfo{author}{\bibfnamefont{D.~D.} \bibnamefont{Solnyshkov}},
  \bibinfo{author}{\bibfnamefont{G.}~\bibnamefont{Malpuech}},
  \bibinfo{author}{\bibfnamefont{G.~B.~H.} \bibnamefont{von H\"ogersthal}},
  \bibinfo{author}{\bibfnamefont{E.}~\bibnamefont{Feltin}},
  \bibnamefont{et~al.}, \bibinfo{journal}{Phys. Rev. Lett.}
  \textbf{\bibinfo{volume}{101}}, \bibinfo{pages}{136409}
  (\bibinfo{year}{2008}).

\bibitem[{\citenamefont{Schneider et~al.}(2013)\citenamefont{Schneider,
  Rahimi-Iman, Kim, Fischer, Savenko, Amthor, Lermer, Wolf, Worschech,
  Kulakovskii et~al.}}]{schneider13a}
\bibinfo{author}{\bibfnamefont{C.}~\bibnamefont{Schneider}},
  \bibinfo{author}{\bibfnamefont{A.}~\bibnamefont{Rahimi-Iman}},
  \bibinfo{author}{\bibfnamefont{N.~Y.} \bibnamefont{Kim}},
  \bibinfo{author}{\bibfnamefont{J.}~\bibnamefont{Fischer}},
  \bibinfo{author}{\bibfnamefont{I.~G.} \bibnamefont{Savenko}},
  \bibinfo{author}{\bibfnamefont{M.}~\bibnamefont{Amthor}},
  \bibinfo{author}{\bibfnamefont{M.}~\bibnamefont{Lermer}},
  \bibinfo{author}{\bibfnamefont{A.}~\bibnamefont{Wolf}},
  \bibinfo{author}{\bibfnamefont{L.}~\bibnamefont{Worschech}},
  \bibinfo{author}{\bibfnamefont{V.~D.} \bibnamefont{Kulakovskii}},
  \bibnamefont{et~al.}, \bibinfo{journal}{Nature}
  \textbf{\bibinfo{volume}{497}}, \bibinfo{pages}{348} (\bibinfo{year}{2013}).

\bibitem[{\citenamefont{Ballarini et~al.}(2013)\citenamefont{Ballarini, Giorgi,
  Cancellieri, Houdr\'e, Giacobino, Cingolani, Bramati, Gigli, and
  Sanvitto}}]{ballarini13a}
\bibinfo{author}{\bibfnamefont{D.}~\bibnamefont{Ballarini}},
  \bibinfo{author}{\bibfnamefont{M.~D.} \bibnamefont{Giorgi}},
  \bibinfo{author}{\bibfnamefont{E.}~\bibnamefont{Cancellieri}},
  \bibinfo{author}{\bibfnamefont{R.}~\bibnamefont{Houdr\'e}},
  \bibinfo{author}{\bibfnamefont{E.}~\bibnamefont{Giacobino}},
  \bibinfo{author}{\bibfnamefont{R.}~\bibnamefont{Cingolani}},
  \bibinfo{author}{\bibfnamefont{A.}~\bibnamefont{Bramati}},
  \bibinfo{author}{\bibfnamefont{G.}~\bibnamefont{Gigli}}, \bibnamefont{and}
  \bibinfo{author}{\bibfnamefont{D.}~\bibnamefont{Sanvitto}},
  \bibinfo{journal}{Nat. Comm.} \textbf{\bibinfo{volume}{4}},
  \bibinfo{pages}{1778} (\bibinfo{year}{2013}).

\bibitem[{\citenamefont{Norris et~al.}(1994)\citenamefont{Norris, Rhee, Sung,
  Arakawa, Nishioka, and Weisbuch}}]{norris94a}
\bibinfo{author}{\bibfnamefont{T.}~\bibnamefont{Norris}},
  \bibinfo{author}{\bibfnamefont{J.-K.} \bibnamefont{Rhee}},
  \bibinfo{author}{\bibfnamefont{C.-Y.} \bibnamefont{Sung}},
  \bibinfo{author}{\bibfnamefont{Y.}~\bibnamefont{Arakawa}},
  \bibinfo{author}{\bibfnamefont{M.}~\bibnamefont{Nishioka}}, \bibnamefont{and}
  \bibinfo{author}{\bibfnamefont{C.}~\bibnamefont{Weisbuch}},
  \bibinfo{journal}{Phys. Rev. B} \textbf{\bibinfo{volume}{50}},
  \bibinfo{pages}{14663} (\bibinfo{year}{1994}).

\bibitem[{\citenamefont{Savona and Weisbuch}(1996)}]{savona96b}
\bibinfo{author}{\bibfnamefont{V.}~\bibnamefont{Savona}} \bibnamefont{and}
  \bibinfo{author}{\bibfnamefont{C.}~\bibnamefont{Weisbuch}},
  \bibinfo{journal}{Phys. Rev. B} \textbf{\bibinfo{volume}{54}},
  \bibinfo{pages}{10835} (\bibinfo{year}{1996}).

\bibitem[{\citenamefont{Wang and Vyas}(1995)}]{wang95a}
\bibinfo{author}{\bibfnamefont{C.}~\bibnamefont{Wang}} \bibnamefont{and}
  \bibinfo{author}{\bibfnamefont{R.}~\bibnamefont{Vyas}},
  \bibinfo{journal}{Phys. Rev. A} \textbf{\bibinfo{volume}{51}},
  \bibinfo{pages}{2516} (\bibinfo{year}{1995}).

\bibitem[{\citenamefont{Marie et~al.}(1999)\citenamefont{Marie, Renucci,
  Dubourg, Amand, Jeune, Barrau, Bloch, and Planel}}]{marie99a}
\bibinfo{author}{\bibfnamefont{X.}~\bibnamefont{Marie}},
  \bibinfo{author}{\bibfnamefont{P.}~\bibnamefont{Renucci}},
  \bibinfo{author}{\bibfnamefont{S.}~\bibnamefont{Dubourg}},
  \bibinfo{author}{\bibfnamefont{T.}~\bibnamefont{Amand}},
  \bibinfo{author}{\bibfnamefont{P.~L.} \bibnamefont{Jeune}},
  \bibinfo{author}{\bibfnamefont{J.}~\bibnamefont{Barrau}},
  \bibinfo{author}{\bibfnamefont{J.}~\bibnamefont{Bloch}}, \bibnamefont{and}
  \bibinfo{author}{\bibfnamefont{R.}~\bibnamefont{Planel}},
  \bibinfo{journal}{Phys. Rev. B} \textbf{\bibinfo{volume}{59}},
  \bibinfo{pages}{2494(R)} (\bibinfo{year}{1999}).

\bibitem[{\citenamefont{Brunetti et~al.}(2006)\citenamefont{Brunetti,
  Vladimirova, Scalbert, Nawrocki, Kavokin, Shelykh, and Bloch}}]{brunetti06a}
\bibinfo{author}{\bibfnamefont{A.}~\bibnamefont{Brunetti}},
  \bibinfo{author}{\bibfnamefont{M.}~\bibnamefont{Vladimirova}},
  \bibinfo{author}{\bibfnamefont{D.}~\bibnamefont{Scalbert}},
  \bibinfo{author}{\bibfnamefont{M.}~\bibnamefont{Nawrocki}},
  \bibinfo{author}{\bibfnamefont{A.~V.} \bibnamefont{Kavokin}},
  \bibinfo{author}{\bibfnamefont{I.~A.} \bibnamefont{Shelykh}},
  \bibnamefont{and} \bibinfo{author}{\bibfnamefont{J.}~\bibnamefont{Bloch}},
  \bibinfo{journal}{Phys. Rev. B} \textbf{\bibinfo{volume}{74}},
  \bibinfo{pages}{241101(R)} (\bibinfo{year}{2006}).

\bibitem[{\citenamefont{Boulier et~al.}(2014)\citenamefont{Boulier, Bamba, Amo,
  Adrados, Lemaitre, Galopin, Sagnes, Bloch, Ciuti, Giacobino
  et~al.}}]{boulier14a}
\bibinfo{author}{\bibfnamefont{T.}~\bibnamefont{Boulier}},
  \bibinfo{author}{\bibfnamefont{M.}~\bibnamefont{Bamba}},
  \bibinfo{author}{\bibfnamefont{A.}~\bibnamefont{Amo}},
  \bibinfo{author}{\bibfnamefont{C.}~\bibnamefont{Adrados}},
  \bibinfo{author}{\bibfnamefont{A.}~\bibnamefont{Lemaitre}},
  \bibinfo{author}{\bibfnamefont{E.}~\bibnamefont{Galopin}},
  \bibinfo{author}{\bibfnamefont{I.}~\bibnamefont{Sagnes}},
  \bibinfo{author}{\bibfnamefont{J.}~\bibnamefont{Bloch}},
  \bibinfo{author}{\bibfnamefont{C.}~\bibnamefont{Ciuti}},
  \bibinfo{author}{\bibfnamefont{E.}~\bibnamefont{Giacobino}},
  \bibnamefont{et~al.}, \bibinfo{journal}{Nat. Comm.}
  \textbf{\bibinfo{volume}{5}} (\bibinfo{year}{2014}).

\bibitem[{\citenamefont{Silva et~al.}(2014)\citenamefont{Silva, {Gonz\'alez
  Tudela}, {S\'anchez Mu\~noz}, Ballarini, Gigli, West, Pfeiffer, del Valle,
  Sanvitto, and Laussy}}]{arXiv_silva14a}
\bibinfo{author}{\bibfnamefont{B.}~\bibnamefont{Silva}},
  \bibinfo{author}{\bibfnamefont{A.}~\bibnamefont{{Gonz\'alez Tudela}}},
  \bibinfo{author}{\bibfnamefont{C.}~\bibnamefont{{S\'anchez Mu\~noz}}},
  \bibinfo{author}{\bibfnamefont{D.}~\bibnamefont{Ballarini}},
  \bibinfo{author}{\bibfnamefont{G.}~\bibnamefont{Gigli}},
  \bibinfo{author}{\bibfnamefont{K.~W.} \bibnamefont{West}},
  \bibinfo{author}{\bibfnamefont{L.}~\bibnamefont{Pfeiffer}},
  \bibinfo{author}{\bibfnamefont{E.}~\bibnamefont{del Valle}},
  \bibinfo{author}{\bibfnamefont{D.}~\bibnamefont{Sanvitto}}, \bibnamefont{and}
  \bibinfo{author}{\bibfnamefont{F.~P.} \bibnamefont{Laussy}},
  \bibinfo{journal}{arXiv:1406.0964}  (\bibinfo{year}{2014}).

\bibitem[{\citenamefont{Dominici et~al.}(2013)\citenamefont{Dominici,
  Ballarini, Giorgi, Cancellieri, Silva, Bramati, Gigli, Laussy, and
  Sanvitto}}]{arXiv_dominici13a}
\bibinfo{author}{\bibfnamefont{L.}~\bibnamefont{Dominici}},
  \bibinfo{author}{\bibfnamefont{D.}~\bibnamefont{Ballarini}},
  \bibinfo{author}{\bibfnamefont{M.~D.} \bibnamefont{Giorgi}},
  \bibinfo{author}{\bibfnamefont{E.}~\bibnamefont{Cancellieri}},
  \bibinfo{author}{\bibfnamefont{B.}~\bibnamefont{Silva}},
  \bibinfo{author}{\bibfnamefont{A.}~\bibnamefont{Bramati}},
  \bibinfo{author}{\bibfnamefont{G.}~\bibnamefont{Gigli}},
  \bibinfo{author}{\bibfnamefont{F.}~\bibnamefont{Laussy}}, \bibnamefont{and}
  \bibinfo{author}{\bibfnamefont{D.}~\bibnamefont{Sanvitto}},
  \bibinfo{journal}{arXiv:1309.3083}  (\bibinfo{year}{2013}).

\bibitem[{\citenamefont{Ciuti and Carusotto}(2005)}]{ciuti05a}
\bibinfo{author}{\bibfnamefont{C.}~\bibnamefont{Ciuti}} \bibnamefont{and}
  \bibinfo{author}{\bibfnamefont{I.}~\bibnamefont{Carusotto}},
  \bibinfo{journal}{Phys. Stat. Sol. B} \textbf{\bibinfo{volume}{242}},
  \bibinfo{pages}{2224} (\bibinfo{year}{2005}).

\bibitem[{sm()}]{sm}
\emph{\bibinfo{title}{See supplementary material.}}

\bibitem[{\citenamefont{Carusotto and Ciuti}(2004)}]{carusotto04a}
\bibinfo{author}{\bibfnamefont{I.}~\bibnamefont{Carusotto}} \bibnamefont{and}
  \bibinfo{author}{\bibfnamefont{C.}~\bibnamefont{Ciuti}},
  \bibinfo{journal}{Phys. Rev. Lett.} \textbf{\bibinfo{volume}{93}},
  \bibinfo{pages}{166401} (\bibinfo{year}{2004}).

\bibitem[{\citenamefont{Cancellieri et~al.}(2010)\citenamefont{Cancellieri,
  Marchetti, Szymanska, and Tejedor}}]{cancellieri10a}
\bibinfo{author}{\bibfnamefont{E.}~\bibnamefont{Cancellieri}},
  \bibinfo{author}{\bibfnamefont{F.~M.} \bibnamefont{Marchetti}},
  \bibinfo{author}{\bibfnamefont{M.~H.} \bibnamefont{Szymanska}},
  \bibnamefont{and} \bibinfo{author}{\bibfnamefont{C.}~\bibnamefont{Tejedor}},
  \bibinfo{journal}{Phys. Rev. B} \textbf{\bibinfo{volume}{82}},
  \bibinfo{pages}{224512} (\bibinfo{year}{2010}).

\bibitem[{\citenamefont{Demirchyan et~al.}(2014)\citenamefont{Demirchyan,
  Chestnov, Alodjants, Glazov, and Kavokin}}]{demirchyan14a}
\bibinfo{author}{\bibfnamefont{S.}~\bibnamefont{Demirchyan}},
  \bibinfo{author}{\bibfnamefont{I.}~\bibnamefont{Chestnov}},
  \bibinfo{author}{\bibfnamefont{A.}~\bibnamefont{Alodjants}},
  \bibinfo{author}{\bibfnamefont{M.}~\bibnamefont{Glazov}}, \bibnamefont{and}
  \bibinfo{author}{\bibfnamefont{A.}~\bibnamefont{Kavokin}},
  \bibinfo{journal}{Phys. Rev. Lett.} \textbf{\bibinfo{volume}{112}},
  \bibinfo{pages}{196403} (\bibinfo{year}{2014}).

\bibitem[{\citenamefont{Laussy et~al.}(2009)\citenamefont{Laussy, del Valle,
  and Tejedor}}]{laussy09a}
\bibinfo{author}{\bibfnamefont{F.~P.} \bibnamefont{Laussy}},
  \bibinfo{author}{\bibfnamefont{E.}~\bibnamefont{del Valle}},
  \bibnamefont{and} \bibinfo{author}{\bibfnamefont{C.}~\bibnamefont{Tejedor}},
  \bibinfo{journal}{Phys. Rev. B} \textbf{\bibinfo{volume}{79}},
  \bibinfo{pages}{235325} (\bibinfo{year}{2009}).

\bibitem[{\citenamefont{Baumberg et~al.}(1998)\citenamefont{Baumberg, Armitage,
  Skolnick, and Roberts}}]{baumberg98a}
\bibinfo{author}{\bibfnamefont{J.~J.} \bibnamefont{Baumberg}},
  \bibinfo{author}{\bibfnamefont{A.}~\bibnamefont{Armitage}},
  \bibinfo{author}{\bibfnamefont{M.~S.} \bibnamefont{Skolnick}},
  \bibnamefont{and} \bibinfo{author}{\bibfnamefont{J.~S.}
  \bibnamefont{Roberts}}, \bibinfo{journal}{Phys. Rev. Lett.}
  \textbf{\bibinfo{volume}{81}}, \bibinfo{pages}{661} (\bibinfo{year}{1998}).

\bibitem[{\citenamefont{Skolnick et~al.}(2000)\citenamefont{Skolnick, Astratov,
  Whittaker, Armitage, Emam-Ismael, Stevenson, Baumberg, Roberts, Lidzey,
  Virgili et~al.}}]{skolnick00a}
\bibinfo{author}{\bibfnamefont{M.~S.} \bibnamefont{Skolnick}},
  \bibinfo{author}{\bibfnamefont{V.~N.} \bibnamefont{Astratov}},
  \bibinfo{author}{\bibfnamefont{D.~M.} \bibnamefont{Whittaker}},
  \bibinfo{author}{\bibfnamefont{A.}~\bibnamefont{Armitage}},
  \bibinfo{author}{\bibfnamefont{M.}~\bibnamefont{Emam-Ismael}},
  \bibinfo{author}{\bibfnamefont{R.~M.} \bibnamefont{Stevenson}},
  \bibinfo{author}{\bibfnamefont{J.~J.} \bibnamefont{Baumberg}},
  \bibinfo{author}{\bibfnamefont{J.~S.} \bibnamefont{Roberts}},
  \bibinfo{author}{\bibfnamefont{D.~G.} \bibnamefont{Lidzey}},
  \bibinfo{author}{\bibfnamefont{T.}~\bibnamefont{Virgili}},
  \bibnamefont{et~al.}, \bibinfo{journal}{J. Lum.}
  \textbf{\bibinfo{volume}{87}}, \bibinfo{pages}{25} (\bibinfo{year}{2000}).

\bibitem[{\citenamefont{Vishnevsky et~al.}(2012)\citenamefont{Vishnevsky,
  Solnyshkov, Gippius, and Malpuech}}]{vishnevsky12a}
\bibinfo{author}{\bibfnamefont{D.~V.} \bibnamefont{Vishnevsky}},
  \bibinfo{author}{\bibfnamefont{D.~D.} \bibnamefont{Solnyshkov}},
  \bibinfo{author}{\bibfnamefont{N.~A.} \bibnamefont{Gippius}},
  \bibnamefont{and} \bibinfo{author}{\bibfnamefont{G.}~\bibnamefont{Malpuech}},
  \bibinfo{journal}{Phys. Rev. B} \textbf{\bibinfo{volume}{85}},
  \bibinfo{pages}{155328} (\bibinfo{year}{2012}).

\bibitem[{\citenamefont{Wouters et~al.}(2013)\citenamefont{Wouters, Para\"iso,
  L\'eger, Cerna, Morier-Genoud, Portella-Oberli, and
  Deveaud-Pl\'edran}}]{wouters13a}
\bibinfo{author}{\bibfnamefont{M.}~\bibnamefont{Wouters}},
  \bibinfo{author}{\bibfnamefont{T.~K.} \bibnamefont{Para\"iso}},
  \bibinfo{author}{\bibfnamefont{Y.}~\bibnamefont{L\'eger}},
  \bibinfo{author}{\bibfnamefont{R.}~\bibnamefont{Cerna}},
  \bibinfo{author}{\bibfnamefont{F.}~\bibnamefont{Morier-Genoud}},
  \bibinfo{author}{\bibfnamefont{M.~T.} \bibnamefont{Portella-Oberli}},
  \bibnamefont{and}
  \bibinfo{author}{\bibfnamefont{B.}~\bibnamefont{Deveaud-Pl\'edran}},
  \bibinfo{journal}{Phys. Rev. B} \textbf{\bibinfo{volume}{87}},
  \bibinfo{pages}{045303} (\bibinfo{year}{2013}).

\end{thebibliography}

\begin{thebibliography}{11}
\expandafter\ifx\csname natexlab\endcsname\relax\def\natexlab#1{#1}\fi
\expandafter\ifx\csname bibnamefont\endcsname\relax
  \def\bibnamefont#1{#1}\fi
\expandafter\ifx\csname bibfnamefont\endcsname\relax
  \def\bibfnamefont#1{#1}\fi
\expandafter\ifx\csname citenamefont\endcsname\relax
  \def\citenamefont#1{#1}\fi
\expandafter\ifx\csname url\endcsname\relax
  \def\url#1{\texttt{#1}}\fi
\expandafter\ifx\csname urlprefix\endcsname\relax\def\urlprefix{URL }\fi
\providecommand{\bibinfo}[2]{#2}
\providecommand{\eprint}[2][]{\url{#2}}

\bibitem[{\citenamefont{Laussy}(2012)}]{S_laussy12a}
\bibinfo{author}{\bibfnamefont{F.}~\bibnamefont{Laussy}},
  \emph{\bibinfo{title}{Exciton-polaritons in microcavities}}
  (\bibinfo{publisher}{Springer Berlin Heidelberg}, \bibinfo{year}{2012}), vol.
  \bibinfo{volume}{172}, chap. \bibinfo{chapter}{1.~Quantum Dynamics of
  Polariton Condensates}, pp. \bibinfo{pages}{1--42}, ISBN
  \bibinfo{isbn}{978-3-642-24186-4}.

\bibitem[{\citenamefont{Colas and Laussy}(2014)}]{S_unpub_colas14a}
\bibinfo{author}{\bibfnamefont{D.}~\bibnamefont{Colas}} \bibnamefont{and}
  \bibinfo{author}{\bibfnamefont{F.}~\bibnamefont{Laussy}},
  \bibinfo{journal}{unpublished}  (\bibinfo{year}{2014}).

\bibitem[{mov({\natexlab{a}})}]{S_movieI}
\emph{\bibinfo{title}{See file \texttt{I-SpaceTimeRabiOscillation.avi}}},
  \bibinfo{howpublished}{Supplementary Movie}.

\bibitem[{\citenamefont{{L\'opez Carre\~no} et~al.}(2014)\citenamefont{{L\'opez
  Carre\~no}, {Restrepo Cuartas}, del Valle, and Laussy}}]{S_unpub_carreno14a}
\bibinfo{author}{\bibfnamefont{J.}~\bibnamefont{{L\'opez Carre\~no}}},
  \bibinfo{author}{\bibfnamefont{J.}~\bibnamefont{{Restrepo Cuartas}}},
  \bibinfo{author}{\bibfnamefont{E.}~\bibnamefont{del Valle}},
  \bibnamefont{and} \bibinfo{author}{\bibfnamefont{F.}~\bibnamefont{Laussy}},
  \bibinfo{journal}{Unpublished}  (\bibinfo{year}{2014}).

\bibitem[{\citenamefont{Demirchyan et~al.}(2014)\citenamefont{Demirchyan,
  Chestnov, Alodjants, Glazov, and Kavokin}}]{S_demirchyan14a}
\bibinfo{author}{\bibfnamefont{S.}~\bibnamefont{Demirchyan}},
  \bibinfo{author}{\bibfnamefont{I.}~\bibnamefont{Chestnov}},
  \bibinfo{author}{\bibfnamefont{A.}~\bibnamefont{Alodjants}},
  \bibinfo{author}{\bibfnamefont{M.}~\bibnamefont{Glazov}}, \bibnamefont{and}
  \bibinfo{author}{\bibfnamefont{A.}~\bibnamefont{Kavokin}},
  \bibinfo{journal}{Phys. Rev. Lett.} \textbf{\bibinfo{volume}{112}},
  \bibinfo{pages}{196403} (\bibinfo{year}{2014}).

\bibitem[{\citenamefont{Spreeuw and Woerdman}(1993)}]{S_spreeuw93a}
\bibinfo{author}{\bibfnamefont{R.}~\bibnamefont{Spreeuw}} \bibnamefont{and}
  \bibinfo{author}{\bibfnamefont{J.}~\bibnamefont{Woerdman}},
  \bibinfo{journal}{Progress in Optics} \textbf{\bibinfo{volume}{31}},
  \bibinfo{pages}{263} (\bibinfo{year}{1993}).

\bibitem[{\citenamefont{del Valle}(2010)}]{S_delvalle_book10a}
\bibinfo{author}{\bibfnamefont{E.}~\bibnamefont{del Valle}},
  \emph{\bibinfo{title}{Microcavity Quantum Electrodynamics}}
  (\bibinfo{publisher}{VDM Verlag}, \bibinfo{year}{2010}).

\bibitem[{\citenamefont{Husimi}(1940)}]{S_husimi40a}
\bibinfo{author}{\bibfnamefont{K.}~\bibnamefont{Husimi}},
  \bibinfo{journal}{Proc. Phys. Math. Soc. Jpn} \textbf{\bibinfo{volume}{22}},
  \bibinfo{pages}{264} (\bibinfo{year}{1940}).

\bibitem[{mov({\natexlab{b}})}]{S_movieII}
\emph{\bibinfo{title}{See file \texttt{II-RabiOscillations.avi}}},
  \bibinfo{howpublished}{Supplementary Movie}.

\bibitem[{mov({\natexlab{c}})}]{S_movieIII}
\emph{\bibinfo{title}{See file \texttt{III-RabiDephasing.avi}}},
  \bibinfo{howpublished}{Supplementary Movie}.

\bibitem[{\citenamefont{Dominici et~al.}(2013)\citenamefont{Dominici,
  Ballarini, Giorgi, Cancellieri, Silva, Bramati, Gigli, Laussy, and
  Sanvitto}}]{S_arXiv_dominici13a}
\bibinfo{author}{\bibfnamefont{L.}~\bibnamefont{Dominici}},
  \bibinfo{author}{\bibfnamefont{D.}~\bibnamefont{Ballarini}},
  \bibinfo{author}{\bibfnamefont{M.~D.} \bibnamefont{Giorgi}},
  \bibinfo{author}{\bibfnamefont{E.}~\bibnamefont{Cancellieri}},
  \bibinfo{author}{\bibfnamefont{B.}~\bibnamefont{Silva}},
  \bibinfo{author}{\bibfnamefont{A.}~\bibnamefont{Bramati}},
  \bibinfo{author}{\bibfnamefont{G.}~\bibnamefont{Gigli}},
  \bibinfo{author}{\bibfnamefont{F.}~\bibnamefont{Laussy}}, \bibnamefont{and}
  \bibinfo{author}{\bibfnamefont{D.}~\bibnamefont{Sanvitto}},
  \bibinfo{journal}{arXiv:1309.3083}  (\bibinfo{year}{2013}).

\end{thebibliography}

\end{document}